\documentclass[a4paper,11pt]{article}
\pdfoutput=1
\usepackage{jheppub}
\usepackage{multirow}

\usepackage{array}
\usepackage{longtable}

\usepackage[usenames,dvipsnames]{xcolor}
\hypersetup{
  colorlinks=true,
  citecolor=blue,
  linkcolor=blue,
  urlcolor=blue}
 
\definecolor{X575}{rgb}{0.05, 0.7, 0.05}

\def \MET{\rm E{\!\!\!/}_T}

\newcommand{\viz}{\it viz.}

\title{Boosting the charged Higgs search prospects using jet substructure at the LHC}

\author{Jinmian Li,$^{a,b}$}
\author{Riley Patrick,$^{a}$}
\author{Pankaj Sharma$^{a}$ and}
\author{Anthony G. Williams$^{a}$}

\affiliation{$^a$Center of Excellence for Particle Physics at Terascale, University of Adelaide, 5005 Adelaide, South Australia}
\affiliation{$^b$School of Physics, Korea Institute for Advanced Study, Seoul 130-722, Korea}

\emailAdd{jinmian.li@adelaide.edu.au}
\emailAdd{riley.patrick@adelaide.edu.au}
\emailAdd{pankaj.sharma@adelaide.edu.au} 
\emailAdd{anthony.williams@adelaide.edu.au}

\abstract{
Charged Higgs bosons are predicted in variety of theoretically well-motivated new physics models with extended Higgs sectors. In this study, we focus on a type-II two Higgs doublet model (2HDM-II) and  consider a heavy charged Higgs with its mass ranging from 500 GeV to 1 TeV as dictated by the $b\to s\gamma$ constraints which render $M_{H^\pm}>480$ GeV. We study the dominant production mode $H^\pm t$ associated production with $H^\pm \to W^\pm A$ being the dominant decay channel when the pseudoscalar $A$ is considerably lighter. For such a heavy charged Higgs, both the decay products $W^\pm$ and $A$ are relatively boosted. In such a scenario, we apply the jet substructure analysis of tagging the fat pseudoscalar and $W$ jets in order to eliminate the standard model background efficiently. We perform a detailed detector simulation for the signal and background processes at the 14 TeV LHC. We introduce various kinematical cuts to determine the signal significance for a number of benchmark points with charged Higgs boson mass from 500 GeV to 1 TeV in the $W^\pm A$ decay channel. Finally we perform a multivariate analysis utilizing a boosted decision tree algorithm to optimize these significances.
}


\begin{document}
 
\maketitle

\section{Introduction}

The Large Hadron Collider (LHC) achieved a milestone when it discovered a 125 GeV scalar boson in its first run \cite{Chatrchyan:2012xdj,Aad:2012tfa}. Even though the current LHC data points to a SM like scalar particle, it is still not sufficient to completely resolve the issue. Through the production and decays of the Higgs to the SM fermions and gauge bosons, the LHC has already measured many of its couplings. However the data as yet allows wide deviations from the SM expectations to within 2$\sigma$. Thus it is still too early to regard the SM as the ultimate theory of particle interactions and there is a need to explore alternative scenarios of electroweak symmetry breaking (EWSB) beyond the SM (BSM) which would be tested in the next run of the LHC, or possibly, at an $e^+e^-$ collider which now has a reasonable hope of being constructed.

There are several theoretically well motivated scenarios to explain the EWSB beyond the minimal SM Higgs sector. One of the simplest and minimal extension of the SM Higgs sector is to include another Higgs doublet. With two Higgs doublets, the scalar sector of the two Higgs doublet model (2HDM) now contains five scalar eigenstates: a light CP even neutral Higgs $h$, a heavy CP even neutral Higgs $H$, a CP odd neutral Higgs $A$ and a pair of charged Higgs $H^\pm$. Depending upon how these two scalar doublets couple with the SM fermions, there can be four distinct 2HDMs, namely, type I, type II, type Y and type X. A recent review of the phenomenology of 2HDM can be found in Ref.~\cite{Branco:2011iw}.

Any signal of a charged Higgs boson at the LHC would be an unambiguous discovery of new physics beyond the SM. However the search for the charged Higgs is quite complicated. If it is lighter than the top quark, it would be profusely produced from the decays of the top quark in top pair production. Such a light $H^\pm$ would dominantly decay to $\tau^\pm \nu$. A detailed study of this decay in top pair and single top productions has been performed in Refs.~\cite{Aoki:2011wd,Guedes:2012eu}. In the high mass region where $M_{H^\pm}>M_t$, the dominant production of $H^\pm$ at the LHC is in association with a single top occurring via the $bg\to tH^-$ + c.c. fusion process \cite{Gunion:1986pe}. However search prospects in the high mass region for $H^\pm$ is quite difficult, owing to large backgrounds to dominant decay process $H^\pm \to tb$. So long as $\tan\beta$, the ratio of the vacuum expectation value of two Higgs doublets, is sufficiently small ($\leq 1.5$) or large ($\geq 30$), the charged Higgs has a reasonable prospects of discovery in $tb$ decay mode \cite{Gunion:1993sv,Barger:1993th,Miller:1999bm,Moretti:1999bw,Dev:2014yca}. A recent exhaustive analysis of the discovery prospects of charged Higgs can be found in Ref.~\cite{Akeroyd:2016ymd}. Searches for charged Higgs in type II 2HDM in associated production with a top quark using the top polarization has been performed in Refs.~\cite{Huitu:2010ad,Godbole:2011vw,Rindani:2013mqa,Gong:2014tqa,Gong:2012ru,Cao:2013ud}.

In this work, among all 2HDM Yukawa types, we mainly focus on the 2HDM model of type II, wherein, the charged Higgs mass is constrained to be larger than 480 GeV according to the $b\to s\gamma$ measurements \cite{Misiak:2015xwa}. Within the framework of the 2HDM, when other neutral scalars such as $H$ and $A$ are lighter, then $H^\pm\to W^\pm\phi$ ($\phi\equiv (h,H,A)$) decays are kinematically open and a new realm in the search for charged Higgs in bosonic decays is available. Recent studies \cite{Moretti:2016jkp,Arhrib:2016wpw,Coleppa:2014cca} have demonstrated the potential of bosonic channels $H^\pm\to W^\pm \phi$  for the $H^\pm$ searches at the LHC. In Ref.~\cite{Moretti:2016jkp}, the inverse alignment scenario of type II 2HDM was considered and their conclusion was that the $W^\pm\phi$ decay mode can be utilized to detect $H^\pm$ in the early run of 14 TeV LHC with only 100 fb$^{-1}$ of integrated luminosity. 

Jet substructure methods in the context of charged Higgs searches have been utilized in Ref.~\cite{Yang:2011jk} where authors have studied a full hadronic decay mode of the charged Higgs and top quark when they are produced in association with each other. They utilized various boosted top-tagging algorithms to reconstruct the boosted hadronic top emanating from the decay of charged Higgs. They concluded that the sensitivity of the LHC to the heavy charged Higgs boson with two $b$ taggings can reach upto 9.5$\sigma$ significance for a 1 TeV charged Higgs. The jet substructure analysis is also found to be useful in probing other Higgs particles in BSM models with extended Higgs sectors~\cite{Kang:2013rj,Chen:2014dma,Hajer:2015gka,Casolino:2015cza,Conte:2016zjp,Goncalves:2016qhh}.

Spurred on by the aforementioned results and the fact that a heavy charged Higgs would lead to boosted decay products, we intend to utilize the jet substructure tools \cite{Butterworth:2008iy} in order to identify boosted Higgs and $W$ bosons. We will then reassess the $H^\pm$ discovery prospects in the $W^\pm\phi$ decay modes in the context of 2HDM at the 14 TeV LHC. We first study a simple cut-based data analysis on the signal and background events. For this, we design a set of kinematical cuts which are found to be suitable for the case of 1 TeV charged Higgs achieving a reasonable signal significance in the early run of 14 TeV LHC. In pursuit of a better signal significance, we also perform a multivariate analysis (MVA) that takes into account the distribution profiles of many kinematical variables. To maximize the signal to background discrimination, we employ a boosted decision tree (BDT) algorithm that enhances classification performance by sequentially boosting the decision trees using the training data.

The plan of the paper is as follows. In the section II, we discuss the production and decays of charged Higgs in 2HDM at the LHC. In section III we discuss the signal and background processes; and various benchmark points for further analysis. In section IV we identify the two signal regions for the analysis, perform a full signal-to-background analysis using jet substructure tools and construct various kinematical variables for the boosted decision tree analysis. Finally we discuss our conclusions from these results.

\section{Production and decay of heavy charged Higgs boson in 2HDM}
\subsection{Production of $H^\pm$}
We consider the charged Higgs production in association with a single top at the LHC. At the parton level, this production occurs via 
\begin{equation}
 g(p_1)b(p_2)\to t(p_3)H^-(p_4).
\end{equation}
There are two Feynmann diagrams contributing to the process: the $s$-channel exchange of a bottom quark and the $t$-channel exchange of a top quark. It is the $H^\pm tb$ coupling in the 2HDM which is relevant for the production cross section. The $H^\pm tb$ coupling in a general 2HDM can be written as
\begin{equation}
 g_{H^\pm t b}=\frac{g}{\sqrt{2}M_W}[m_t \mathcal A_t P_L + m_b \mathcal A_b P_R],
\end{equation}
where $g$ is the SU(2) gauge coupling and where $\mathcal A_t$ and $\mathcal A_b$ have been defined in Table~\ref{tab:At} for type I and II 2HDMs.

\begin{table}[h!]\centering
 \begin{tabular}{|c|c|c|}\hline
  Models & $\mathcal A_t$ & $\mathcal A_b$\\\hline\hline
  type I & $1/\tan\beta$  & $1/\tan\beta$\\
  type II & $1/\tan\beta$  & $\tan\beta$\\\hline
 \end{tabular}
 \caption{\label{tab:At} The values of $\mathcal A_t$ and $\mathcal A_b$ for type I and II 2HDMs.}
\end{table}

\begin{figure}[h!]\centering
 \includegraphics[scale=0.6]{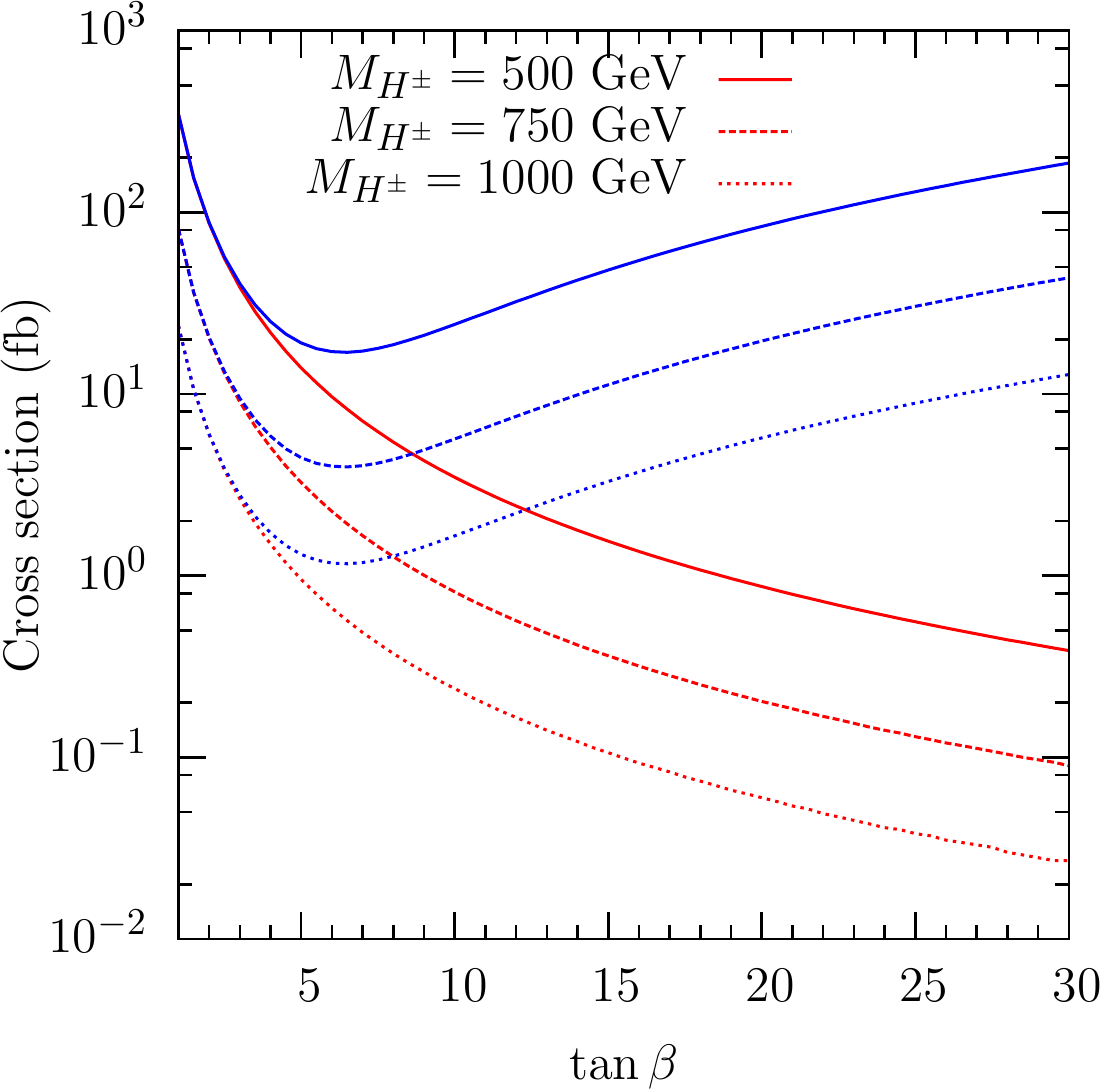}\hspace{10mm}
 \includegraphics[scale=0.6]{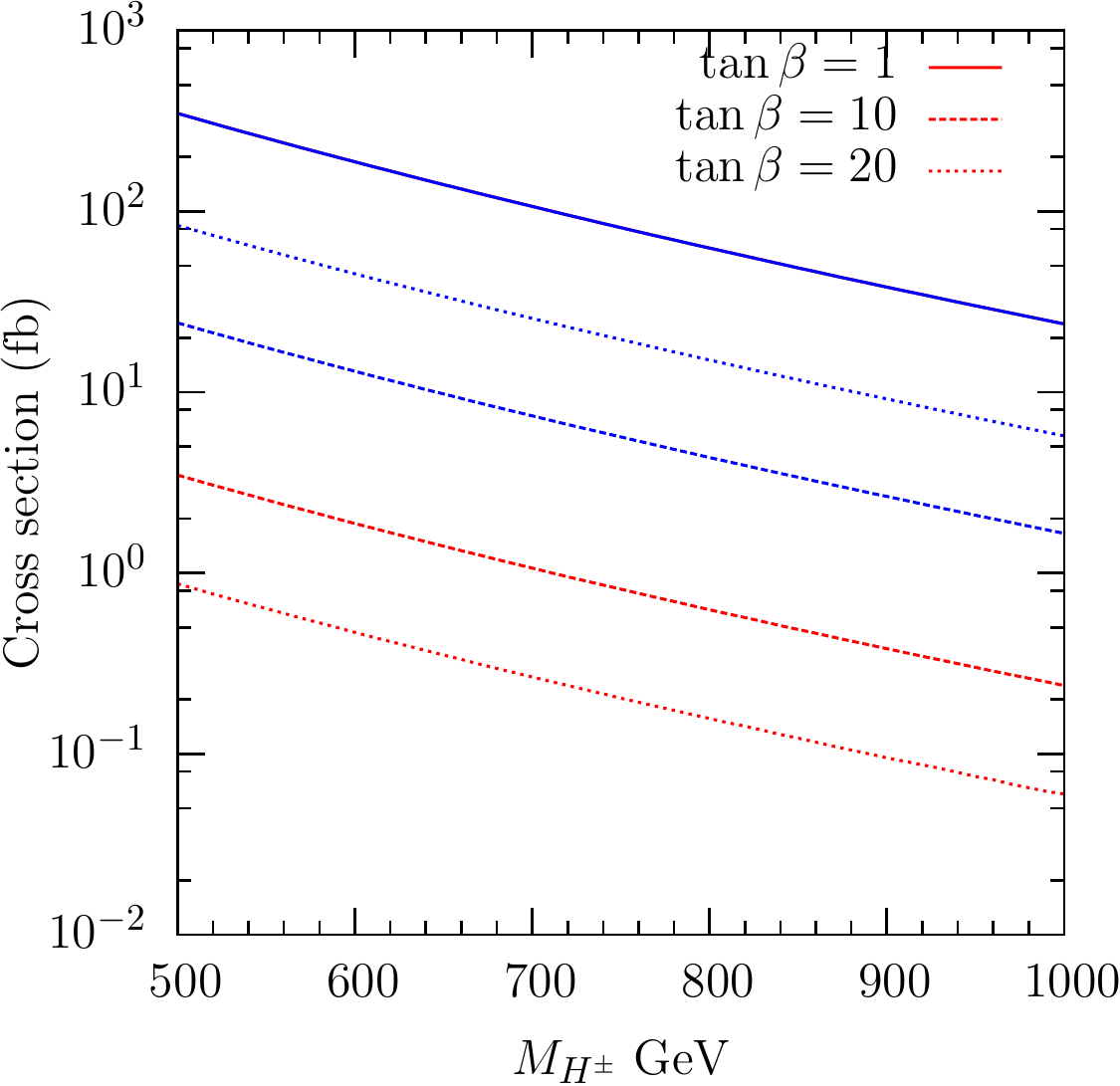}
 \caption{\label{fig:xsec}Cross section for inclusive $tH^-$ production at the 14 TeV LHC as a function of $\tan\beta$ (left) and mass of the charged Higgs $M_{H^\pm}$. The red and blue color curves denote the cross sections for type I and type II 2HDMs. Note that in right panel, the red and blue solid curves overlap each other for $\tan\beta=1$.}
\end{figure}

The production cross section for the $tH^-$ process is proportional to $[m_t^2~\mathcal A_t^2 + m_b^2~\mathcal A_b^2]$ in a general 2HDM. For a type I 2HDM, as both $\mathcal A_t$ and $\mathcal A_b$ are inversely proportional to $\tan\beta$, the production cross section decreases rapidly with increasing $\tan\beta$. For a type II 2HDM, first the cross section goes down with increasing $\tan\beta$ until reaching its minimum value at $\tan\beta=\sqrt{m_t/m_b}\sim 6.4$. This is independent of center-of-mass energy and charged Higgs mass. As $\tan\beta$ is increased further the $m_b^2\tan\beta^2$ term takes over leading to rise in production cross section. This behaviour is seen in the Fig.~\ref{fig:xsec}, where the cross section as a functions of $\tan\beta$ (left panel) and $H^\pm$ mass (right panel) has been displayed for type I and II 2HDMs. At large values, $\tan\beta\geq 30$, the production cross section in the type I 2HDM becomes almost insignificant while for type II case it is as significant as for the lower $\tan\beta$ values. In the figure, the red (blue) curves denote the cross section for type I (II) 2HDM. In the right panel of the figure, for $\tan\beta=1$, the cross sections for type I and II models are equal and thus the two corresponding curves overlap with each other.

\subsection{Decays of $H^\pm$}

The tree level decay modes for a heavy charged Higgs relevant to our analysis include $H^\pm\to (tb,W^\pm h, W^\pm H, W^\pm A)$ with $h$, $H$ and $A$ being the light CP even neutral scalar, the heavy CP even neutral scalar and CP odd scalar respectively. The partial decay widths for each channel can be expressed as 

\begin{subequations}
\begin{eqnarray}
\Gamma[H^\pm\to tb]&=& \frac{3M_{H^\pm}}{8\pi v^2}[m_t^2 \mathcal A_t^2 + m_b^2 \mathcal A_b^2]\sqrt{1-4r_t},\\ 
\Gamma[H^\pm\to W^\pm h]&=& \frac{g^2 c_{\beta-\alpha}^2}{64\pi M_{H^\pm}} \lambda^{1/2}\left(1,r_W,r_h\right)\nonumber\\
                         &\times& \left[m_W^2-2(M_{H^\pm}^2+M_h^2)+\frac{(M_{H^\pm}^2-M_h)^2}{m_W^2}\right],\\
\Gamma[H^\pm\to W^\pm H]&=& \frac{g^2 s_{\beta-\alpha}^2}{64\pi M_{H^\pm}} \lambda^{1/2}\left(1,r_W,r_H\right)\nonumber\\
                         &\times& \left[m_W^2-2(M_{H^\pm}^2+M_H^2)+\frac{(M_{H^\pm}^2-M_H)^2}{m_W^2}\right],\\
\Gamma[H^\pm\to W^\pm A]&=& \frac{g^2 }{64\pi M_{H^\pm}} \lambda^{1/2}\left(1,r_W,r_A\right)\nonumber\\
                         &\times& \left[m_W^2-2(M_{H^\pm}^2+M_A^2)+\frac{(M_{H^\pm}^2-M_A)^2}{m_W^2}\right],                         
\end{eqnarray}
\end{subequations}
where $r_X=m_X^2/M_{H^\pm}^2$, $c_{\beta-\alpha}\equiv \cos(\beta-\alpha)$, $s_{\beta-\alpha}\equiv \sin(\beta-\alpha)$ and $\lambda^{1/2}(1,x,y)\equiv[(1-x^2-y^2)^2-4x^2y^2]^{1/2}$.

\begin{figure}[t!]\centering
 \includegraphics[width=7.5cm]{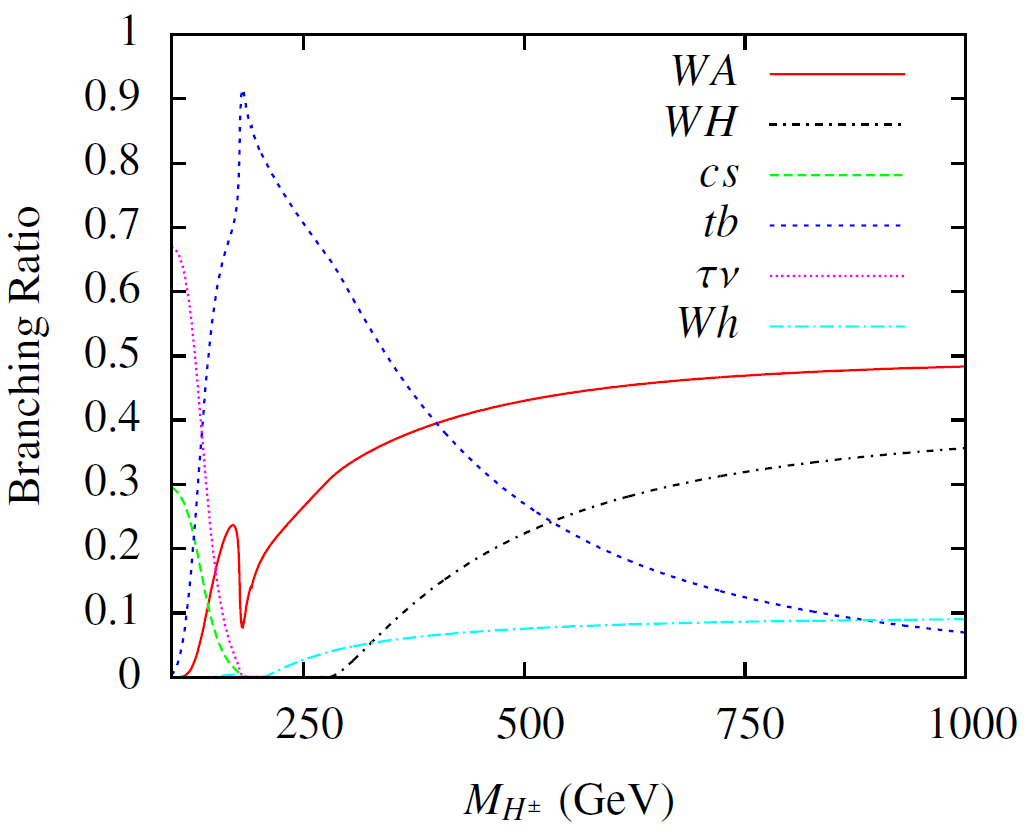}
 \includegraphics[width=7.45cm]{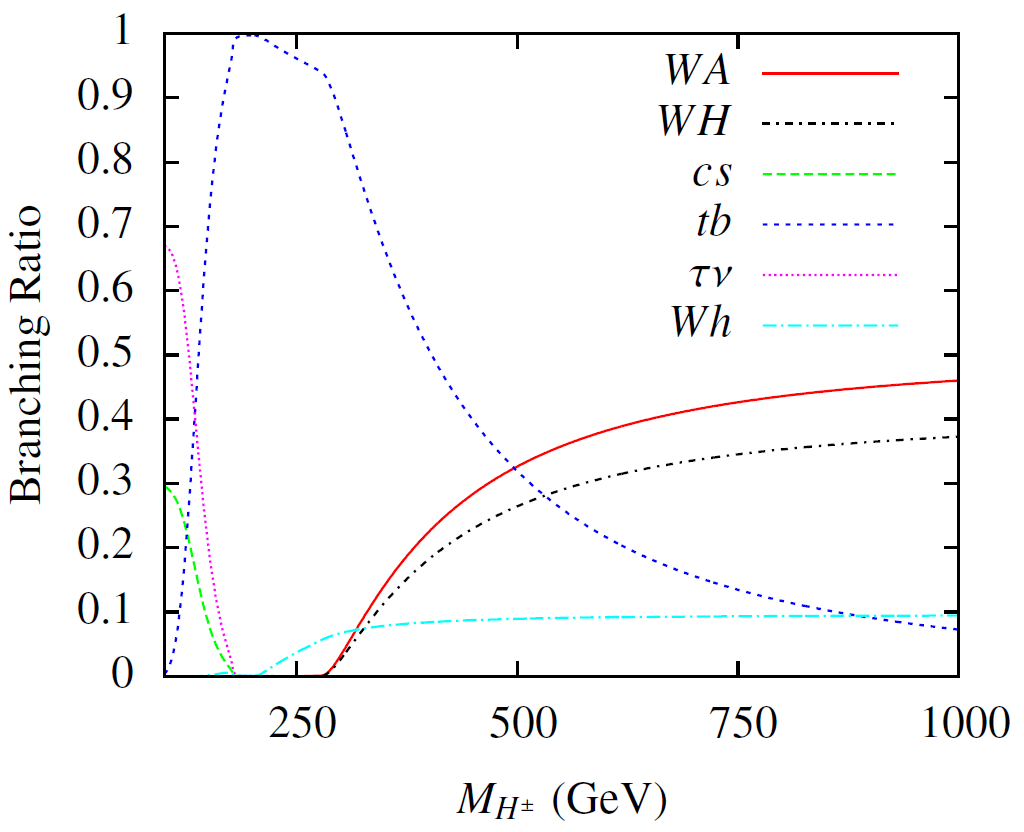}
 \caption{\label{fig:br_Hc} Branching ratios of the $H^\pm$ to different decay channels as a function of its mass for two values of pseudoscalar masses $100$ GeV (left ) and 200 GeV (right). The values of $\tan\beta$ and $s_{\beta-\alpha}$ are chosen to be 1 and 0.9 respectively. We also choose heavy CP even Higgs mass to be 200 GeV in both the plots.}
\end{figure}

\begin{figure}[h!]\centering
\includegraphics[scale=0.35]{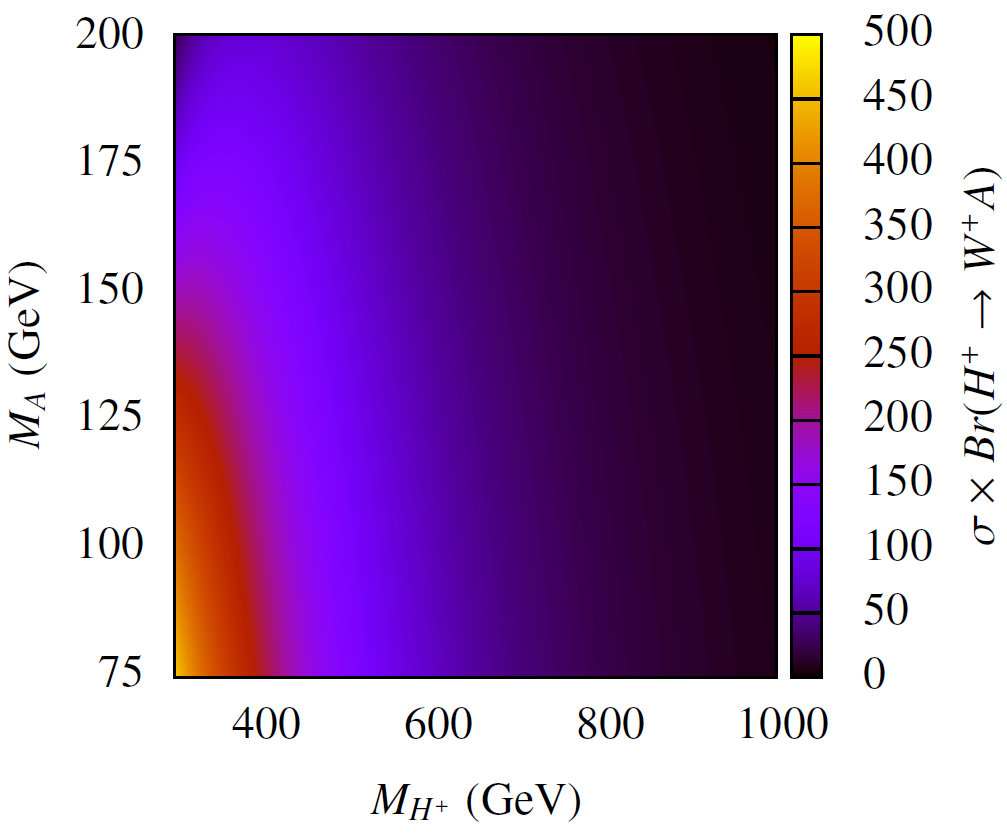}
\caption{\label{fig:xsec_mA}Cross section ($\sigma(pp\to tH^-)$ + c.c) times BR($H^\pm\to W^\pm A$) in the plane of ($M_{H^\pm},M_A$) at 14 TeV LHC. The color bar denotes the resulting cross section in fb.}
\end{figure}

The decay width of $H^\pm\to tb$ depends on the parameters $\tan\beta$ and $M_{H^\pm}$. For a type I 2HDM, with increasing $\tan\beta$, the decay branching ratio (BR) of the $tb$ decay mode goes down as $m_t^2/\tan^2\beta$. On the other hand, for the type II 2HDM, it first decreases until $\tan\beta=\sqrt{m_t/m_b}$ then rises significantly as $m_b^2\tan^2\beta$ as it does for for the $tH^-$ production cross section. The bosonic decays of $H^\pm$ to the CP even Higgs bosons ($H^\pm\to W^\pm h$ and $H^\pm \to W^\pm H$) are proportional to the mixing angle $(\beta-\alpha)$ between $h$ and $H$. The BR for the former decay is proportional to $c_{\beta-\alpha}^2$ while the latter to $s_{\beta-\alpha}^2$. The current LHC data prefers the alignment limits of 2HDM, i.e., the $s_{\beta-\alpha}\sim 1$ ($c_{\beta-\alpha}\sim 1$) for the case when $h$ ($H$) is the light SM-like Higgs. In such a scenario, it is easy to see that a charged Higgs couples very weakly to the SM-like Higgs boson. Thus, unfortunately $H^\pm$ searches in the bosonic decays cannot exploit the invariant mass reconstruction of SM-like scalar around 125 GeV. On the other hand, the $H^\pm$ decay in the $W^\pm A$ channel is not suppressed by any mixing angle and thus can be dominant over other decays if $A$ is light enough. Moreover, in the alignment limit of the 2HDM, the BR of the $W^\pm A$ decay mode becomes equal to that of the $W^\pm H$ mode if $A$ and $H$ are degenerate in mass. 

For the purpose of illustration, in Fig.~\ref{fig:br_Hc} we show the branching ratios of the charged Higgs to various decay channels, namely, $tb$, $W^\pm h$, $W^\pm H$ and $W^\pm A$ for $s_{\beta-\alpha}=0.9$ and $\tan\beta=1$ in type II 2HDM. We show the variation of these BRs with the mass of the charged Higgs $H^\pm$ from 100 GeV until 1 TeV. We have chosen the $m_h=125$ GeV, $m_H=200$ GeV and $m_A=100$ GeV (200 GeV) in the left (right) panels of the Fig. In the work, we mainly focus on the bosonic decays of the charged Higgs to pseudoscalar though the same analysis can be extended to heavy CP even Higgs. In Fig.~\ref{fig:xsec_mA}, we display the product of the charged Higgs production cross section ($\sigma(pp\to tH^-)$) and the branching ratio of the $H^\pm$ decay to $W^\pm A$ in the plane of the charged Higgs mass and the pseudoscalar mass for $\tan\beta=1$. All the decay widths and branching ratios of charged and pseudoscalar Higgs are obtained using {\tt 2HDMC} \cite{Eriksson:2009ws}.

\section{LHC searches for heavy $H^\pm$ and light $A$}
\subsection{Signature and Backgrounds}

We study a heavy charged Higgs production in association with a top quark at the 14 TeV LHC followed by the decay $H^\pm\to W^\pm A$ with both $W^\pm$ and $A$ being highly boosted when $A$ is light. Thus the final state contains two $W$ bosons coming from the top and $H^\pm$ decays, 3 $b$ jets coming from $A$ and top decays. In this analysis, we consider one of the $W$'s decaying leptonically and the other hadronically leading to a signal $\ell^\pm b\bar b bj j\MET+X$. The largest background for the signal comes from the $WWbbj$ process which includes the top pair production plus one jet process. It mimics the signal when the light jet is mistagged as a $b$ jet. The irreducible background comes from the $WWbbb$ process which includes the top pair production associated  with a $b$ quark. Another background considered in this analysis comes from the $WWbjj$ process, which has significant cross section but can be manageable using $b$ identification.

In order to generate the signal and background events at leading order, we use {\tt Madgraph5} \cite{Alwall:2014hca}. Further, we use \texttt {PYTHIA8.2} \cite{Sjostrand:2007gs} to perform parton showers and hadronization for both signal and background events. All events are then passed to {\tt DELPHES3} \cite{deFavereau:2013fsa} for the fast detector simulation, where we apply the default  ATLAS detector card. The \texttt {DELPHES3} output is then used for jet substructure analysis using \texttt {FastJet}\cite{Cacciari:2011ma}.

\subsection{Benchmark points for the analysis}

The choice of benchmark points in this analysis is dictated by the fact that jet substructure methods work best in the scenarios where the mass difference between the charged Higgs and the pseudoscalar boson is large. Thus to demonstrate the utility and limitation of the jet substructure analysis we choose benchmark points with various mass differences between $H^\pm$ and $A$. We consider three values of the charged Higgs mass, $M_H^\pm=500$ GeV, 750 GeV and 1 TeV and three values of the CP odd Higgs mass, $M_A=$ 100 GeV, 150 GeV and 200 GeV. Thus, in total, we study 9 benchmark points: BP1: (500, 100) GeV, BP2: (500, 150) GeV, BP3: (500, 200) GeV, BP4: (750, 100) GeV, BP5: (750, 150) GeV, BP6: (750, 200) GeV, BP7: (1000, 100) GeV, BP8: (1000, 150) GeV and BP9: (1000, 200) GeV. 

We take $M_h=125$ GeV and $\tan\beta=1$ in the analysis. As this analysis does not depend on the CP property of the neutral scalars that $H^\pm$ decays to, it is equally applicable to signals in which $H^\pm\to W^\pm h/H$. As mentioned in the foregoing, the current LHC data prefers the alignment scenario leading to almost equal coupling of the pseudoscalar and the heavy CP even scalar to charged Higgs. Thus including the contribution of $H$ into our analysis may further improve the signal cross section and in turn achieve a better signal-to-background ratio.

\begin{figure}[b!]\centering
 \includegraphics[scale=0.44]{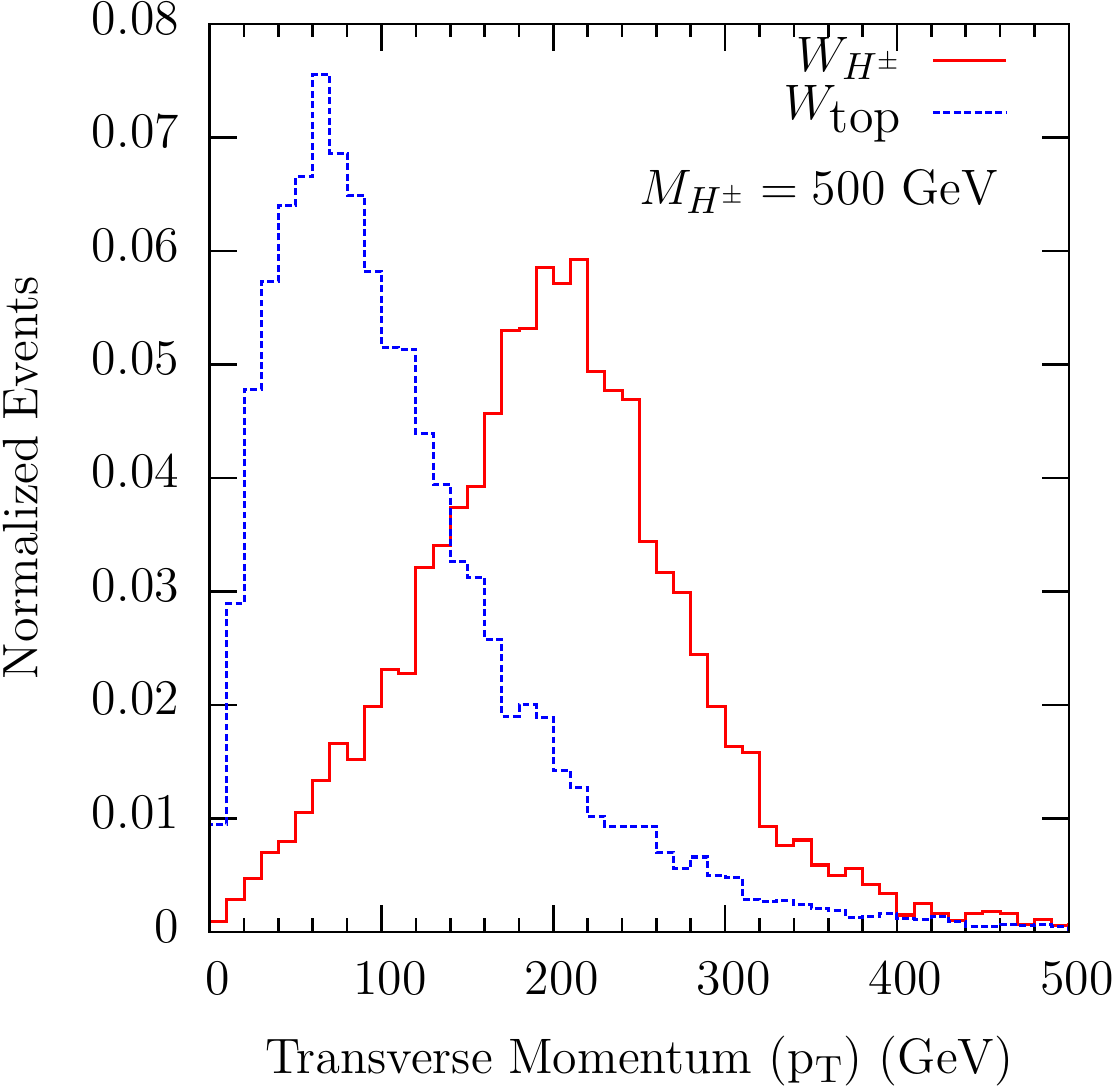}
 \includegraphics[scale=0.44]{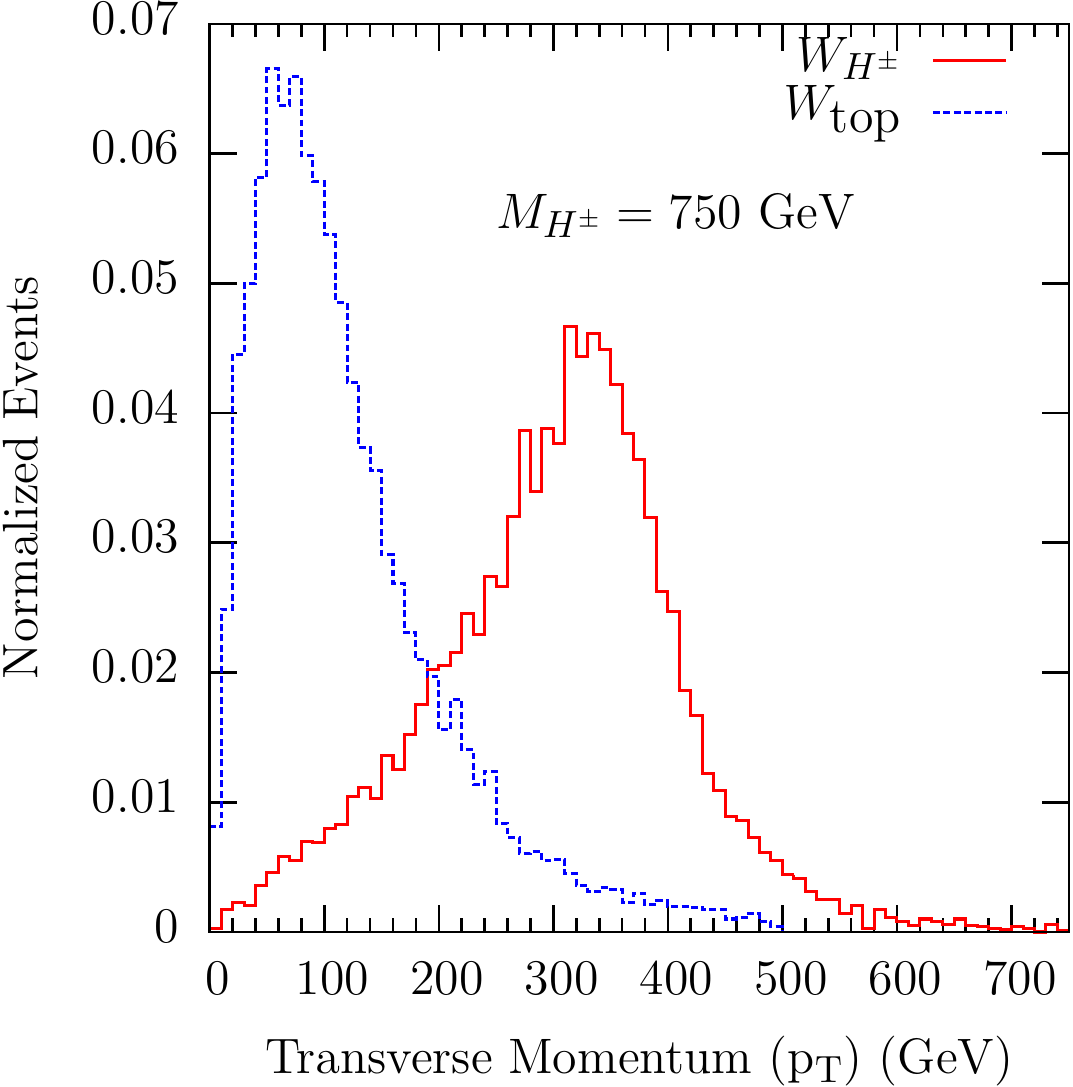}
 \includegraphics[scale=0.44]{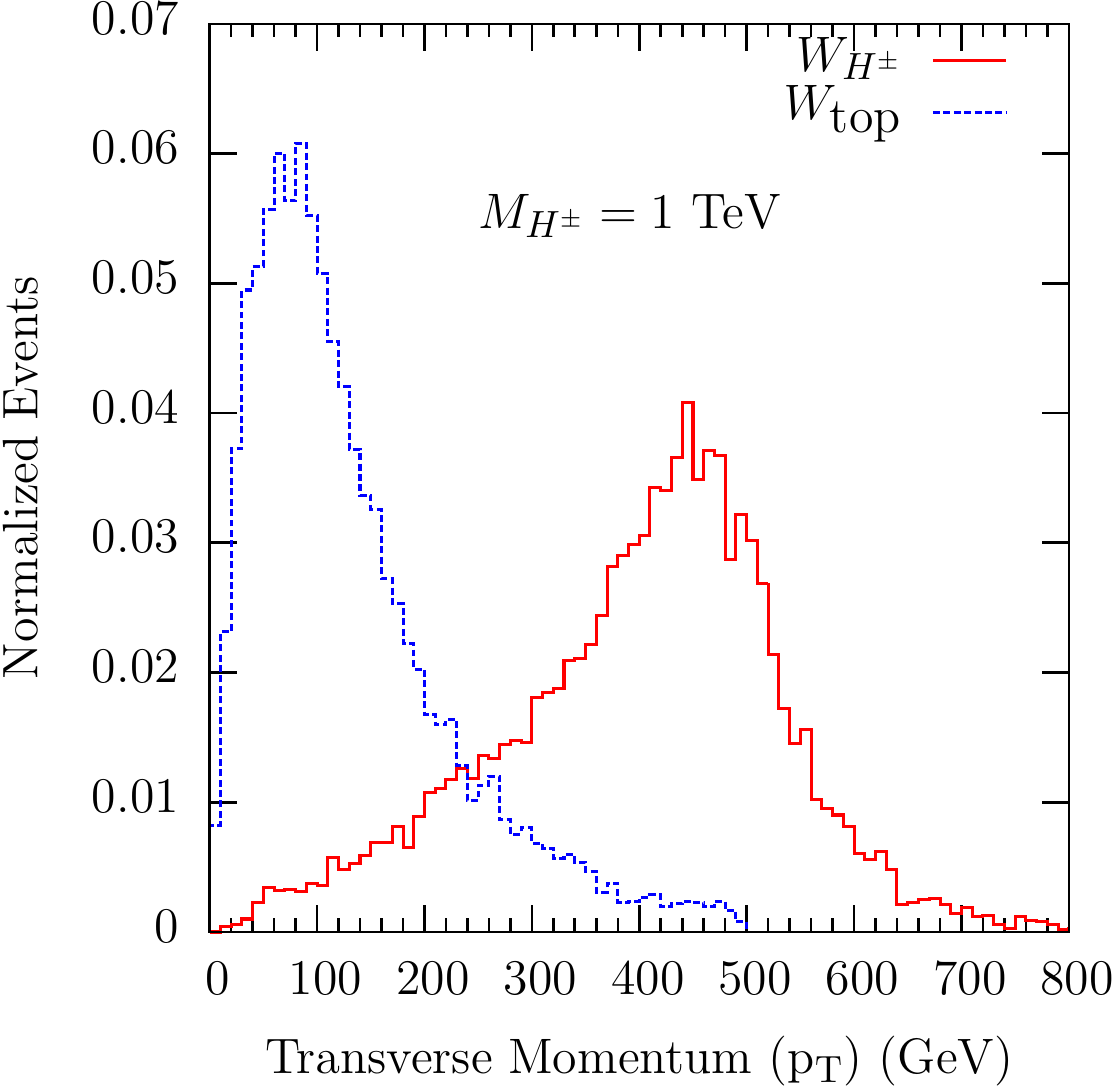}
\caption{\label{fig:pT_W}Transverse momentum of the $W$ boson coming from the top decay vs the one coming from the charged Higgs decays. The plots for three different values of $H^\pm$ masses 500 GeV (left), 750 GeV (middle) and 1 TeV (right) have been shown.}
 \end{figure}
 \begin{figure}[h!]\centering
\includegraphics[scale=0.75]{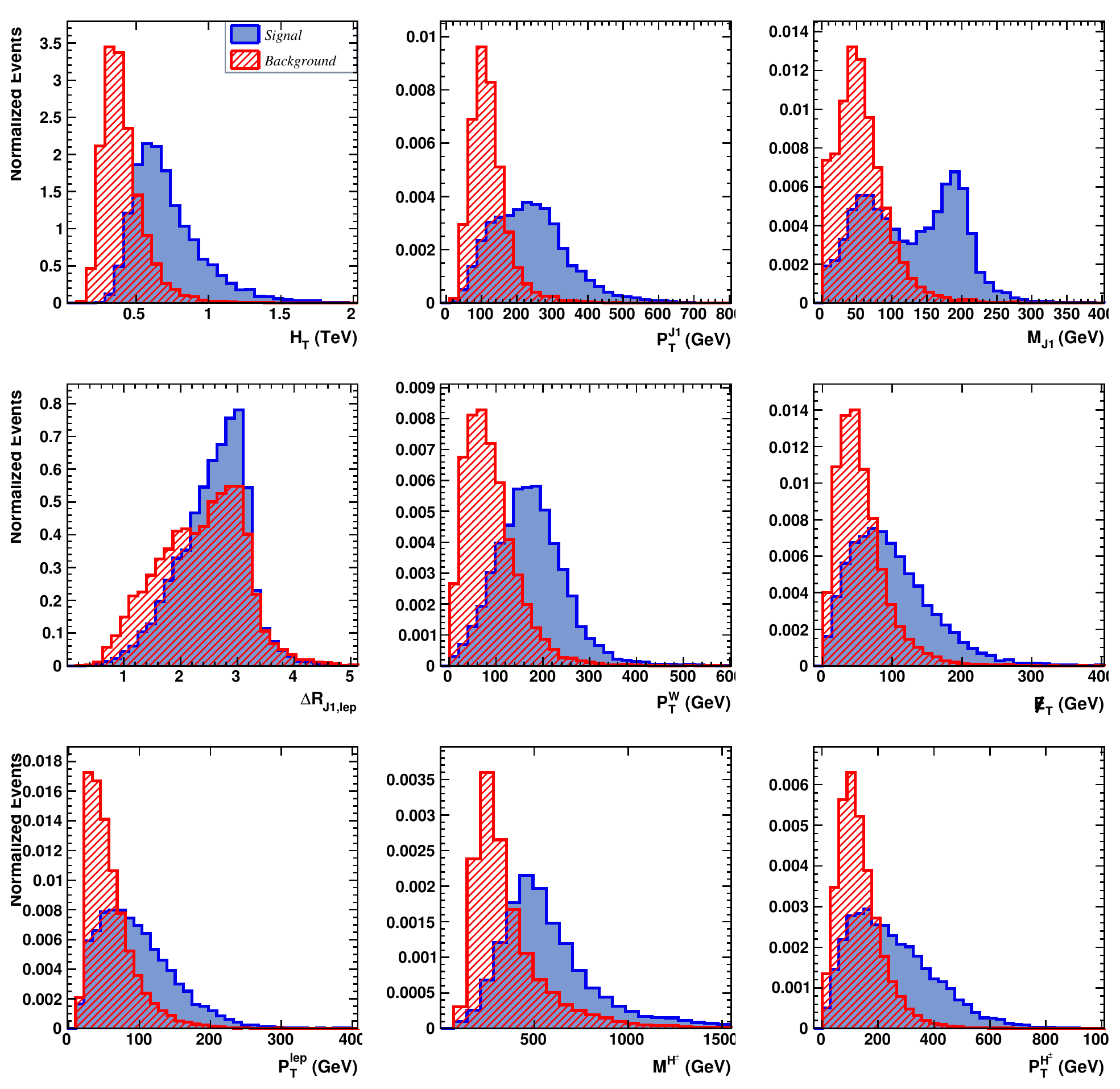} 
\caption{\label{fig:kin_var_bdt_I_3}Kinematic variables for the cut-based and multivariate analysis in BDT for SRI for BP3.}
\end{figure}

\section{Analysis}
\subsection{Framework}
Our search strategy heavily relies on the very boosted Higgs boson which opportunely enhances the signal-to-background ratio. We start our analysis with the preselection of the objects. The particle-flow charged tracks, after isolating the charged leptons, the particle-flow neutral hadrons, and the particle-flow photons in the {\tt DELPHES3} output are used for jet reconstruction. The fat jets are clustered using the Cambridge-Aachen (CA) jet algorithm with a particular jet cone size of $R=1.2$ in order to capture all the collimated decay products of boosted bosons. We then apply the BDRS algorithm which utilizes the mass-drop technique to identify the substructure inside a reconstructed fat jet. This is followed by filtering in which we recluster the constituents of fat jets with radius  $R_{\text{filt}}$ $= \text{min}(0.35; R_{12}/2)$ and select the three hardest subjects to suppress the pileup effects\footnote{ In ref. \cite{Krohn:2013lba} it has been found that the jet grooming techniques such as filtering are quite effective in suppressing pileup and underlying events. Though we do not include these events in simulation, we do perform the filtering in order to include its effect on signal and background effects. Note however that including the pileup and underlying events into our simulation may worsen the significance attained in the analysis.}. 

The Higgs identification then requires the tagging of two $b$ jets among the three filtered subjets. We assume 70\% $b$ tagging efficiency with a 1\% mistagging rate for light flavor and gluon jets \cite{Aad:2015ydr}. To tag the $b$ jets inside the boosted Higgs jet, both ATLAS \cite{ATLAS:2016wlr} and CMS \cite{CMS:2013vea} use ``subjet b tagger'' that first identifies two subjets within the Higgs jet and then applies the standard $b$ tagging algorithm with the similar efficiency as that of an isolated $b$ jet to each subjet. In \cite{CMS:2016jdj} CMS  proposed the ``double-b tagger'' method which first identifies the displaced vertices within the Higgs jet, and then combines information from these vertices with other jet quantities in a dedicated multivariate algorithm. This method shows a better tagging efficiency than the ``subjet b tagger'' (see fig. 3 of ref.\cite{CMS:2016jdj}). In our analysis, we use ``subjet b tagger'' which both CMS and ATLAS have used in their analysis of boosted Higgs as mentioned earlier.

After the Higgs tag is successful, we remove the constituents of the corresponding fat jet from the event and recluster the remaining remnants in the events using the anti-k$_T$ jet clustering algorithm with jet cone radius of $ R=0.4$.

\begin{figure}[h!]\centering
\includegraphics[scale=0.75]{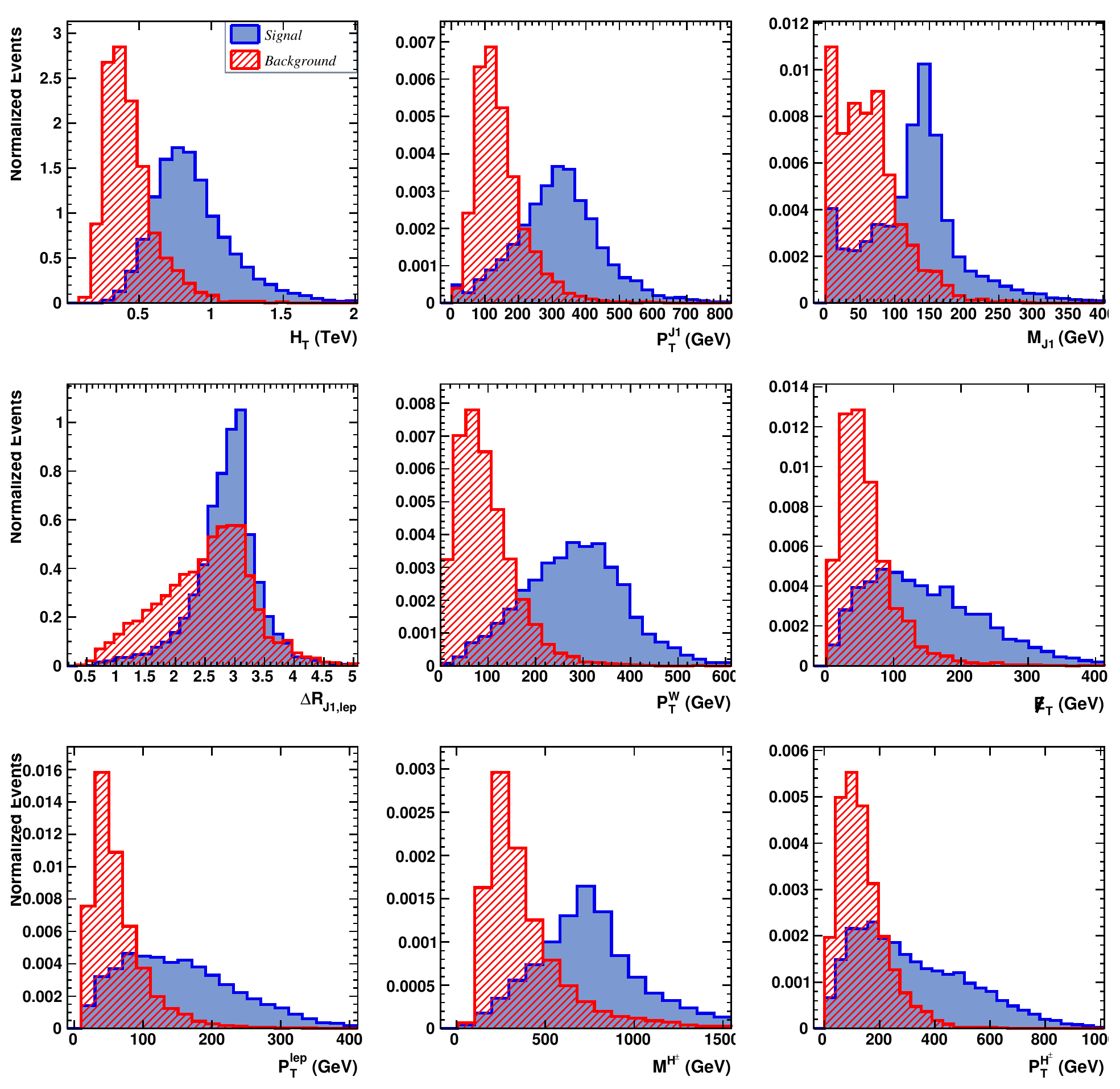} \vspace{15pt}
\caption{\label{fig:kin_var_bdt_I_5}Kinematic variables for the cut-based and multivariate analysis in BDT for SRI for BP5.}
\end{figure}

\begin{figure}[h!]\centering
\includegraphics[scale=0.75]{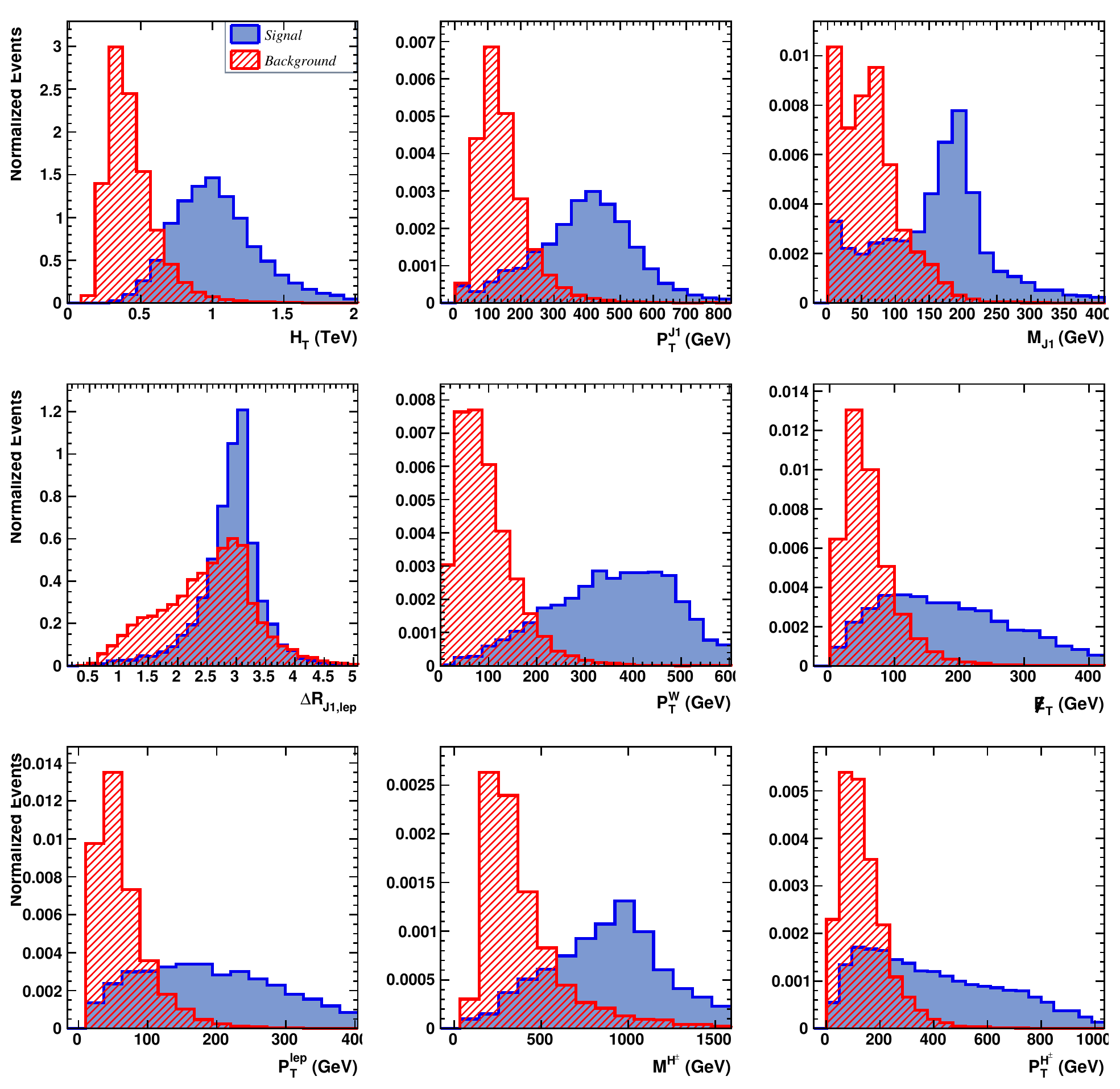} \vspace{15pt}
\caption{\label{fig:kin_var_bdt_I_9}Kinematic variables for the cut-based and multivariate analysis in BDT for SRI for BP9.}
\end{figure}

In what follows, we define two signal regions, aiming for two different decay modes: (1) one where the top decays hadronically while the charged Higgs decays semileptonically (SRI) and (2) vice versa (SRII). In SRII, the leptonic $W$ ($W_{\text{lep}}$) would come from the top and thus is likely to be relatively softer with low $p_T$ while in SRI, as it comes from a heavy charged Higgs, it is likely to be quite harder. This fact can be seen in Fig.~\ref{fig:pT_W}. In SRI the hadronic $W$ boson is reconstructed via two light narrow jets, while for SRII decay products of the hadronic $W$ boson are collimated along its direction and thus it appears as a fat jet whose invariant mass shows a peak around $M_W$. In both signal regions the longitudinal component of the neutrino momentum ($p_{\nu L}$) coming from $W_{\text{lep}}$ can be determined, using the information of the missing transverse momentum and by imposing the invariant mass constraint $M_{\ell\nu} = M_{W^\pm}$, as
\begin{equation}
p_{\nu L}=\frac{1}{2p_{\ell T}^2}\left(A_W p_{\ell L} \pm E_\ell \sqrt{A_W^2\pm 4 p_{\ell T}^2 E_{\nu T}^2}\right)
\end{equation}
where $A_W=M_W^2+2\vec{p}_T\cdot \vec{E}_{\nu T}$. In the case where there are two solutions, we adopt the one which gives $M_{\ell\nu}$ closer to the $W$-mass. We reject events with complex solutions. The four momentum of $W_{\text{lep}}$ is obtained by the vector sum of the charged lepton and neutrino momenta. The $p_T$ of the leptonic $W$ boson is used to separate the two signal regions kinematically i.e., an event is attributed to SRI if $p_T(W_{\text{lep}})>$ 150, 200, 250 GeV for $M_{H^\pm}=$ 500, 750 and 1 TeV respectively and attribute to SRII otherwise.

\begin{figure}[h!]\centering
\includegraphics[scale=0.75]{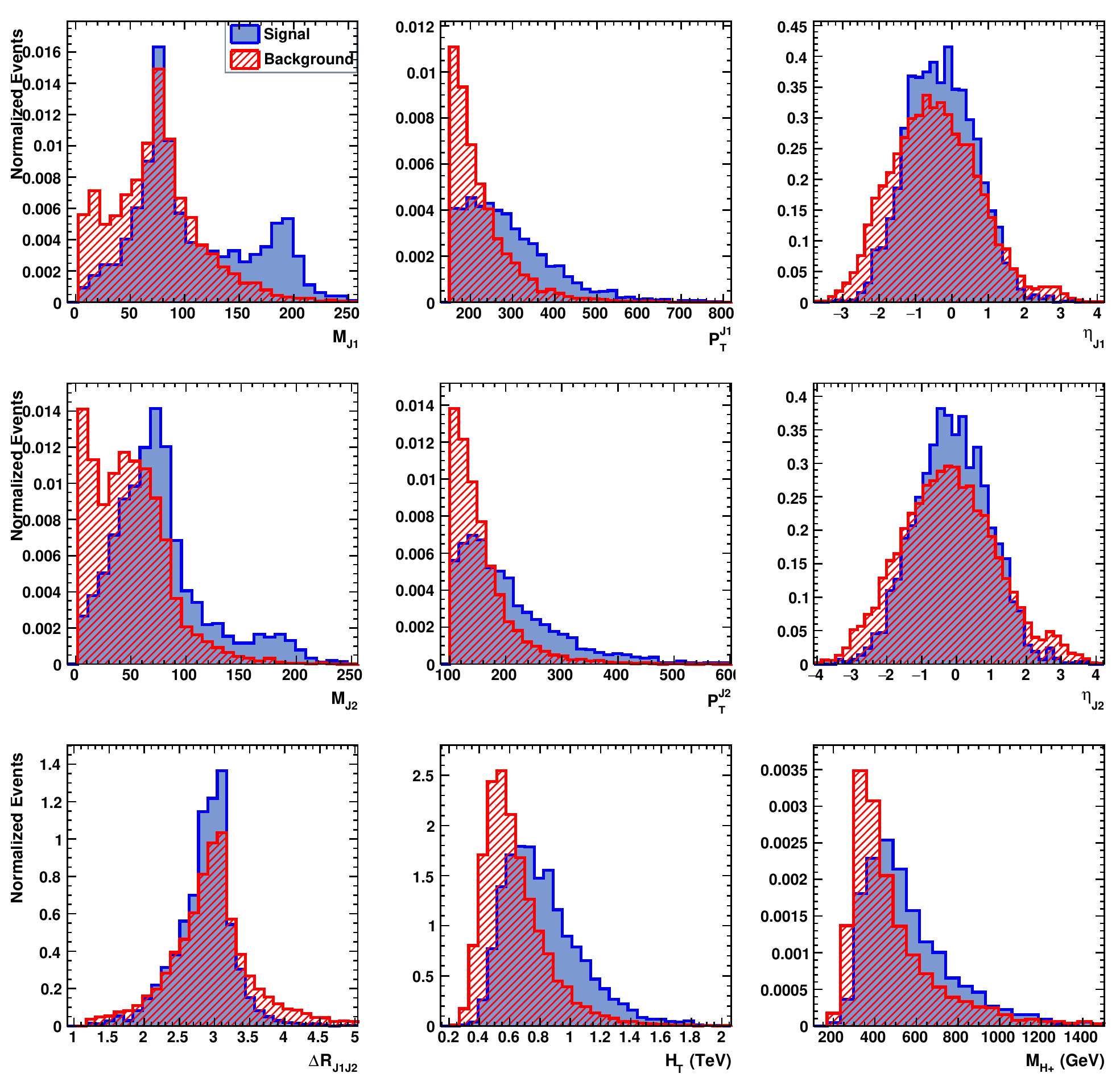} \vspace{15pt}
\caption{\label{fig:kin_var_bdt_II_3}Kinematic variables for the cut-based and multivariate analysis in BDT for SRII for BP3.}
\end{figure}

\begin{figure}[h!]\centering
\includegraphics[scale=0.75]{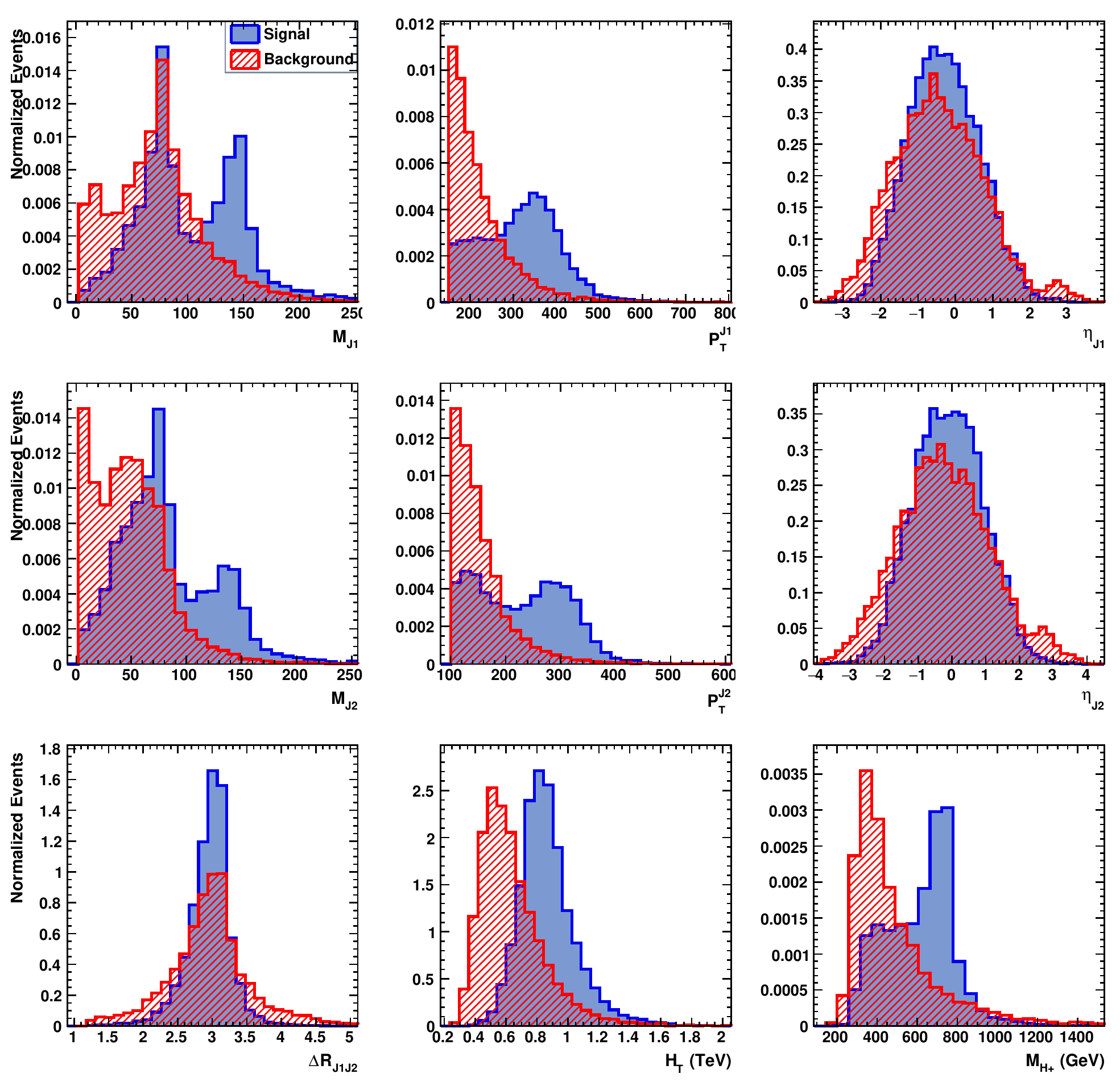} \vspace{15pt}
\caption{\label{fig:kin_var_bdt_II_5}Kinematic variables for the cut-based and multivariate analysis in BDT for SRII for BP5.}
\end{figure}

\begin{figure}[h!]\centering
\includegraphics[scale=0.75]{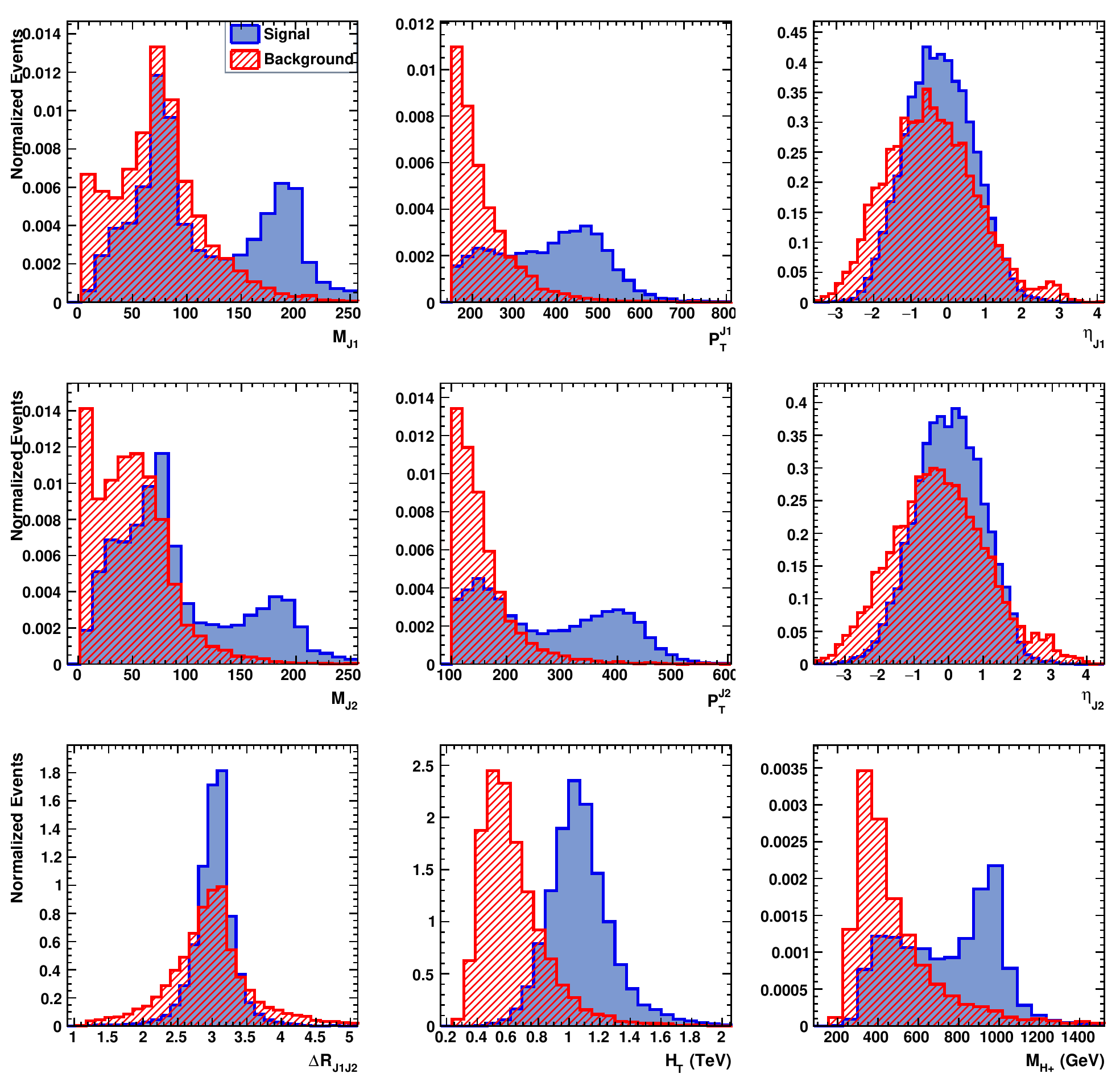} \vspace{15pt}
\caption{\label{fig:kin_var_bdt_II_9}Kinematic variables for the cut-based and multivariate analysis in BDT for SRII for BP9.}
\end{figure}

\subsection{Cut-based Analysis}
For the sake of comparison with the multivariate analysis that we perform in the next section, we study the signal efficiency with respect to background after applying a series of well optimized cuts. Before describing the cuts in detail, we study the various kinematical distributions in each of the signal regions SRI and SRII that can discriminate the signal and background. In the signal region SRI we study the distributions of the following kinematic variables: (1) $H_T$, which is the scalar sum of the transverse momenta of all the visible particles in the final state, $H_T=p_T^{\ell^\pm}+\sum_j p_T^j$; (2) $p_T$ distribution of the leading fat jet, (3) $p_T$ distribution of the leptonic $W$ boson; (4) $p_T$ distribution of the charged lepton; (5) invariant mass distribution of the Higgs jet; (6) missing transverse energy $\MET$; (7) $\Delta$R separation between the Higgs jet and charged lepton; (8) the reconstructed mass of the charged Higgs; and (9) $p_T$ of the reconstructed charged Higgs. We have displayed these distributions in Figs. \ref{fig:kin_var_bdt_I_3}, \ref{fig:kin_var_bdt_I_5} and \ref{fig:kin_var_bdt_I_9} for benchmark points BP3, BP5 and BP9 respectively for region SRI. 

Similarly, for signal region SRII, which exhibits different kinematical characteristics, we study following variables: (1) $H_T$ distribution; (2) invariant mass of the first leading fat jet; (3) invariant mass of the second leading fat jet; (4) $p_T$ distribution of the first leading fat jet; (5) $p_T$ distribution of the second leading fat jet; (6) pseudorapidity distribution of the first leading fat jet; (7) pseudorapidity distributions of the second leading fat jet; (8) $\Delta R$ separation between the first two leading fat jets; and (9) the mass of the charged Higgs reconstructed from the leading two fat jets (one of which must be a Higgs jet). These distributions have been shown in the Figs.~\ref{fig:kin_var_bdt_II_3}, \ref{fig:kin_var_bdt_II_5} and \ref{fig:kin_var_bdt_II_9} for benchmark points BP3, BP5 and BP9 respectively for signal region SRII.

\renewcommand{\arraystretch}{1.1}

\begin{table}[b!]
\begin{center}
\newcolumntype{C}[1]{>{\centering\let\newline\\\arraybackslash\hspace{0pt}}m{#1}}
\begin{tabular}{ ||C{5.8cm}||C{1.65cm}| C{1.65cm} | C{1.75cm} | C{1.65cm} ||}
\hline\hline                                                                  
& \textbf{Signal} & \textbf{WWbbj} & \textbf{WWbbb} & \textbf{WWbjj} \\ \hline
\textbf{{Cross section x BR (fb)}} & 35.0           & 1.6$\times$10$^5$           & 2.6$\times$10$^3$            & 9.8$\times$10$^3$             \\\hline\hline
Trigger                                 	& 0.50         & 0.40         & 0.32         & 0.13  \\\hline
$p_T$ of $W^{lep}$ $\geq$ 150 GeV       	& 0.32         & 6.5$\times$10$^{-2}$        & 4.1$\times$10$^{-2}$         & 2.9$\times$10$^{-2}$ \\\hline
$H_T$ $\geq$ 500 GeV                    	& 0.26         & 3.1$\times$10$^{-2}$         & 1.7$\times$10$^{-2}$        & 1.4$\times$10$^{-2}$ \\\hline
One Higgs Jet 					& 4.1$\times$10$^{-2}$         & 4.0$\times$10$^{-4}$       & 6.0$\times$10$^{-4}$       & 1.0$\times$10$^{-5}$\\\hline
Higgs Jet $p_T\geq$ 200 GeV             	& 3.3$\times$10$^{-2}$         & 2.0$\times$10$^{-4}$       & 3.0$\times$10$^{-4}$        & $\sim$ 0.0 \\\hline
$\MET \geq$ 100 			& 1.6$\times$10$^{-2}$ 	& 6.0$\times$10$^{-5}$ 	& 2.0$\times$10$^{-4}$ 	& $\sim$ 0.0 \\\hline
$|M_{J1\ell\nu} - M_{H^\pm}| \leq 100$ GeV 	& 7.0$\times$10$^{-3}$ 	& 1.0$\times$10$^{-5}$ 	& 3.0$\times$10$^{-5}$	& $\sim$ 0.0 \\\hline\hline
\textbf{Cross section after cuts (fb)}  	& 0.25  &  $\sim$ 1.6        & 7.8$\times$10$^{-2}$          & $\sim$ 0.0          \\
\hline\hline
S/$\sqrt{S+B}$ @ 100fb$^{-1}$                                                      & 1.8               \\ \cline{1-2}
S/$\sqrt{S+B}$ @ 500fb$^{-1}$                                                      & 4.1               \\
\cline{1-2}
\cline{1-2}        
\end{tabular}
\caption{Cut flow of the efficiencies for signal and backgrounds at the 14 TeV LHC in SRI for $M_{H^\pm}=500$ TeV (BP3). \label{tab:cut_eff_I_3}}
\end{center}
\end{table}

After analyzing the kinematical distributions, we devise a set of cuts in each signal regions. Below we list the cuts which we imposed in the signal region SRI:

\begin{enumerate}
\item {\bf Trigger:} Trigger includes all the detector acceptance cuts, namely, $$p_T^{\ell^\pm}>20~ \text{GeV},~ |\eta_{\ell^\pm}|<2.5,~ p_T^{j,b}>20~\text{GeV},~ |\eta_{j,b}|<2.5.$$
In addition, we also impose that there must be exactly one charged lepton in each event. The trigger efficiency is around 40-45\% for the signal as well as background.

\item {\bf A cut on transverse momentum of the leptonic $W$ boson:} To separate the signal region SRI from SRII, we further require the $p_T$ of reconstructed $W_{\text{lep}}^\pm$ to be greater than 150 (250) GeV for charged Higgs of mass of 500 (1000) GeV. 

\item {\bf A cut on $H_T$ distribution:} The $H_T$ distributions for the signal in the case of the heavy charged Higgs are much more harder than the background distributions. It is obvious that for heavy charged Higgs, a stringent $H_T$ cut can be quite detrimental to the backgrounds and thus can be effective in enhancing the signal-to-background ratio. To utilize this fact, we adopt the cut: $H_T> 500(700)$ GeV for $M_{H^\pm}=500(1000)$ GeV.

\item {\bf One Higgs jet:} The leading fat jet in an event must be tagged as a Higgs jet. This step requires $b$ taggings on the leading two filtered subjets inside the fat jet. It turns out that this particular cut is the most effective in suppressing the background. For the signal, the Higgs tagging efficiency is much higher for the 1 TeV $H^\pm$ i.e., 50\% while it is only 10\% for 500 GeV charged Higgs \footnote{For a Higgs with $p_T$ between 400 GeV to 800 GeV, ATLAS \cite{ATLAS:2016wlr} have found the Higgs tagging efficiency to be around 40\%-50\%. For a highly boosted Higgs, the efficiency drops sharply as it becomes increasingly difficult to resolve a fat jet into subjets.}. 

\item {\bf A cut on $p_T$ of the Higgs jet:} As the Higgs is emanated from the decay of a heavy charged Higgs, it is expected to have large transverse momentum. To make use of this, we further impose a cut on its $pT>$ 200 GeV. This cut diminishes the background to half while the signal events are only reduced by 15\%.

\item {\bf Missing transverse energy:} In signal region SRI, the elusive neutrino comes from a heavy $H^\pm$ and thus is expected to carry large $\MET$ while for the background the $\MET$ is quite small. We choose events with $\MET$ larger than 100 GeV.

\item {\bf Charged Higgs Mass Window:} The charged Higgs is reconstructed from the leptonic $W$ and the Higgs jet in SRI. We choose the mass window of the reconstructed charged Higgs according to its real mass. For $M_{H^\pm}=500$ GeV (1 TeV), we select events if the invariant mass of the Higgs jet ($J_1$) and leptonic $W$ boson, $|M_{J_1\ell\nu}-M_{H^\pm}|<100 (200)$ GeV. With this constraint the background events are reduced to half while the signal is almost unchanged.
\end{enumerate}

\begin{table}[b!]
\begin{center}
\newcolumntype{C}[1]{>{\centering\let\newline\\\arraybackslash\hspace{0pt}}m{#1}}
\begin{tabular}{ ||C{5.8cm}||C{1.65cm}| C{1.65cm} | C{1.75cm} | C{1.65cm} ||}
\hline\hline
& \textbf{Signal} & \textbf{WWbbj} & \textbf{WWbbb} & \textbf{WWbjj} \\ \hline
\textbf{{Cross section x BR (fb)}} 	& 5           	& 1.6$\times$10$^5$           	& 2.6$\times$10$^3$            & 9.8$\times$10$^3$             \\\hline\hline
Trigger                                	& 0.51         & 0.40         & 0.32 	& 0.13 \\\hline
$p_T$ of $W^{lep}$ $\geq$ 200 GeV      	& 0.48         & 6.5$\times$10$^{-2}$         & 4.1$\times$10$^{-2}$         & 2.9$\times$10$^{-2}$ \\\hline
$H_T$ $\geq$ 700 GeV                   	& 0.43         & 1.1$\times$10$^{-2}$         & 6.0$\times$10$^{-3}$         & 1.4$\times$10$^{-2}$ \\\hline
One Higgs Jet 				& 0.21         & 8.0$\times$10$^{-5}$       & 1.0$\times$10$^{-4}$        & $\sim$ 0.0 \\\hline
Higgs Jet ${p_T}$ $\geq$ 200 GeV        & 0.21         & 3.0$\times$10$^{-5}$       & 1.0$\times$10$^{-4}$        & $\sim$ 0.0 \\\hline
$\MET \geq$ 100 GeV 		& 0.16 	& 2.0$\times$10$^{-5}$ 	& 6.0$\times$10$^{-5}$ 	& $\sim$ 0.0 \\\hline
$|M_{J1\ell\nu} - M_{H^\pm}| \leq 200$ GeV 	& 8.7$\times$10$^{-2}$ 	& 1.0$\times$10$^{-5}$ 	& 3.0$\times$10$^{-5}$ 	& $\sim$ 0.0 \\\hline\hline
\textbf{Cross section after cuts (fb)}  & 0.44  	&  1.6        	& 7.8$\times$10$^{-2}$         & $\sim$ 0.0 \\\hline\hline
S/$\sqrt{S+B}$ @ 100fb$^{-1}$                           & 3.1                \\ \cline{1-2}

S/$\sqrt{S+B}$ @ 500fb$^{-1}$                            & 6.7                \\
\cline{1-2}
\end{tabular}
\caption{Cut flow of the efficiencies for signal and backgrounds at the 14 TeV LHC in SRI for $M_{H^\pm}=1000$ TeV (BP9).  \label{tab:cut_eff_I_9}}
\end{center}
\end{table}

In Table~\ref{tab:cut_eff_I_3}, we present the cut flow of the efficiencies for the signal for BP3 in signal region SRI and for the different backgrounds. For the BP3 the signal cross section after multiplying the branching ratios of $H^\pm\to W^\pm A$ and $A\to b\bar b$ and before applying any kinematical cuts is 35 fb (including the contribution of the conjugate process). The corresponding total background is 170 pb (after including BR of the two $W$ bosons in the process). We see from the table that the background is reduced to 6.5\% after applying the trigger and the cut on the transverse momentum of the leptonic $W$ boson in the process while the signal events are only reduced to 32\%. This is expected as the $W^\pm_{\text{lep}}$ bosons in the signal region SRI are expected to carry large transverse momentum. The other important cuts happen to be those on $H_T$ and the requirement of at least one Higgs jet in an event which suppress the total background contribution to $\mathcal O(10^{-4})$ of its initial value while the signal events are at 4.1\%. Subsequent cuts on $p_T$ of the Higgs jet, missing transverse energy and mass window around the reconstructed charged Higgs further reduce the signal cross section to  0.25 fb and the final total background cross section turns out to be 1.7 fb. Thus we find the signal significance, $S/\sqrt{S+B}$ for this benchmark point to be  4.1 with integrated luminosity of 500 fb$^{-1}$.  Thus, even the most difficult scenario in our analysis has the reasonable prospects of discovery with around 1 ab$^{-1}$ of data.

The situation improves further for the benchmark point BP9 even though it has a very small production cross section for a charged Higgs of 1 TeV mass. In Table~\ref{tab:cut_eff_I_9}, we show a table for the cut efficiencies and signal efficiencies for the benchmark point BP9. The initial cross section for the signal for this benchmark point is only 5 fb considerably smaller than 35 fb in the benchmark point BP3. However, because of the large signal and background separation in for this mass of $H^\pm$, the signal efficiency comes out to be 8.7\% which is remarkably better than 0.7\% in the benchmark point BP3. This result leads to a much larger significance for BP9 and this benchmark point is within the reach of early LHC data in its 14 TeV run.

\begin{table}[b!]
\begin{center}
 \newcolumntype{C}[1]{>{\centering\let\newline\\\arraybackslash\hspace{0pt}}m{#1}} 
\begin{tabular}{ ||C{5.8cm}||C{1.65cm}| C{1.65cm} | C{1.75cm} | C{1.65cm} ||}\hline \hline
&\textbf{Signal} & \textbf{WWbbj}  & \textbf{WWbbb}  & \textbf{WWbjj}    \\\cline{1-5}
\textbf{{Cross section x BR (fb)}}	& 35.0	& 1.6$\times 10^5$	& 2.6$\times 10^3$	& 9.8$\times 10^3$ 	\\ \hline\hline
{Trigger }				& 0.43	& 0.42			& 0.44			& 0.45 	\\ \hline 
{$p_T$ of $W^{\text{lep}}<150$ GeV }	& 0.33	& 0.32			& 0.33			& 0.34 	\\ \hline
{$H_T>500$ GeV }			& 0.19	& 4.9$\times 10^{-2}$	& 4.3$\times 10^{-2}$	& 4.0$\times 10^{-2}$ 	\\ \hline
{One Higgs jet}				& 1.5$\times 10^{-2}$	& 7.5$\times 10^{-4}$	& 1.3$\times 10^{-4}$	& 1.2$\times 10^{-5}$ \\ \hline
{1st leading fat jet $p_T>150$ GeV}	& 1.2$\times 10^{-2}$	& 3.5$\times 10^{-4}$	& 7.0$\times 10^{-5}$	& 5.8$\times 10^{-6}$ \\ \hline
{2nd leading fat jet $p_T>100$ GeV}	& 1.0$\times 10^{-2}$	& 1.4$\times 10^{-4}$	& 4.0$\times 10^{-5}$	& 2.2$\times 10^{-6}$ \\ \hline
{$ |M_{J_1J_2}-M_{H^\pm}|< 100 \ \mbox{GeV}$}& 6.0$\times 10^{-3}$	& 4.0$\times 10^{-5}$	& 1.2$\times 10^{-5}$	& $\sim 0.0$ \\ \hline
\hline
\textbf{Cross section after cuts (fb)}	 & 0.21 & 6.4 & 0.03 & $\sim$ 0.0\\\hline\hline
$S/\sqrt{S+B}$ @ 1000 fb$^{-1}$		& 	2.6			\\\cline{1-2}
$S/\sqrt{S+B}$ @ 3000 fb$^{-1}$		& 	4.5			\\\cline{1-2}
\end{tabular}
\caption{ Cut flow of the efficiencies for signal and backgrounds at the 14 TeV LHC in SRII for $M_{H^\pm}=500$ GeV (BP3). \label{tab:cut_eff_II_3}}
\end{center}
\end{table}

Below we list the various cuts which we applied in the SRII:

\begin{enumerate}
\item {\bf Trigger:} Trigger includes all the detector acceptance cuts, namely, $$p_T^{\ell^\pm}>20~ \text{GeV},~ |\eta_{\ell^\pm}|<2.5,~ p_T^{j,b}>20~\text{GeV},~ |\eta_{j,b}|<2.5.$$
In addition, we also require that there must be exactly one charged lepton in each event. The trigger efficiency is around 40-45\% for the signal as well as background.

\item {\bf A cut on transverse momentum of the leptonic $W$ boson:} To separate the signal region SRII from SRI, we require the $p_T$ of reconstructed $W_{\text{lep}}^\pm$ to be smaller than 150 (250) GeV for a charged Higgs of mass of 500 (1000) GeV. 

\item {\bf A cut on $H_T$ distribution:} This cut is the same as in signal region SRI.

\item {\bf One Higgs jet:} One of the two leading fat jets in an event must be tagged as a Higgs jet. As mentioned earlier, this requires tagging the two subjets inside the fat jet as $b$ jets. In SRII, we find that the Higgs tagging efficiency for the signal is 50 (10)\% for 1 TeV (500 GeV) $H^\pm$. 

\item {\bf Cuts on $p_T$'s of first two leading fat jets:} The decay of a heavy charged Higgs in the signal leads to two fat jets with the high $p_T$'s while for the background, these are expected to be soft. To utilize this fact we impose cuts on  $p_T^{}J_1>$ 150 GeV and $p_T^{}J_1>$ 100 GeV. 

\item {\bf Charged Higgs Mass Window:} The charged Higgs in signal region SRII is reconstructed from the two hardest fat jets, one of which must be tagged as a Higgs jet. As earlier, for $M_{H^\pm}=500$ GeV (1 TeV), we select events if the invariant mass of the two hardest fat jets ($J_1$ and $J_2$), $|M_{J1J2}-M_{H^\pm}|<100 (200)$ GeV. 
\end{enumerate}

We now discuss the effects of kinematical cuts on the benchmark points BP3 and BP9 in the signal region SRII. The cut flow efficiencies of the signal and various backgrounds are presented in Tables~\ref{tab:cut_eff_II_3} and \ref{tab:cut_eff_II_9}. Unlike in the signal region SRI where the cut on the $p_T$ of the leptonic $W$ boson is quite stringent and suppress the total background by a factor of $\sim$ 15, it is not much effective in the signal region SRII. However in the signal region SRII there is another cut that we find to be much effective {\viz.} a cut on the $p_T$ of the 2nd leading fat jet which suppresses the background contribution by an order of magnitude for the benchmark point BP9. Moreover the reconstructed charged Higgs invariant mass distribution is quite separated for BP9 than for BP3. All these facts result in significantly better suppression of the total background in the benchmark point BP9 than for BP3. Consequently, the search prospects are far better for a 1 TeV charged Higgs in BP9 than for a 500 GeV $H^\pm$ in BP3 in the early run of the 14 TeV LHC.

\begin{table}[b!]
\begin{center}
 \newcolumntype{C}[1]{>{\centering\let\newline\\\arraybackslash\hspace{0pt}}m{#1}} 
\begin{tabular}{ ||C{5.8cm}||C{1.65cm}| C{1.65cm} | C{1.75cm} | C{1.65cm} ||}\hline \hline
&\textbf{Signal} & \textbf{WWbbj}  & \textbf{WWbbb}  & \textbf{WWbjj}    \\\cline{1-5}
\textbf{{Cross section x BR (fb)}}	& 5.0		& 1.6$\times 10^5$	& 2.6$\times 10^3$	& 9.8$\times 10^3$ \\ \hline\hline
{Trigger }				& 0.43		& 0.42			& 0.44			& 0.45 	\\ \hline 
{$p_T$ of $W^{\text{lep}}<250$ GeV }	& 0.41		& 0.40			& 0.42			& 0.38 	\\ \hline
{$H_T>800$ GeV }			& 0.30		& 1.3$\times 10^{-2}$	& 9.2$\times 10^{-3}$	& 1.6$\times 10^{-2}$ 	\\ \hline
{One Higgs jet}				& 0.19		& 7.5$\times 10^{-4}$	& 1.1$\times 10^{-4}$	& 6.7$\times 10^{-6}$ \\ \hline
{1st leading fat jet $p_T>150$ GeV}	& 0.17		& 2.5$\times 10^{-4}$	& 7.2$\times 10^{-5}$	& $\sim 0.0$ \\ \hline
{2nd leading fat jet $p_T>100$ GeV}	& 0.16		& 1.4$\times 10^{-5}$	& 9.0$\times 10^{-6}$	& $\sim 0.0$ \\ \hline
{$ |M_{J1J2}-M_{H^\pm}|< 200 \ \mbox{GeV}$}& 0.10		& 9.3$\times 10^{-6}$	& 4.2$\times 10^{-6}$	& $\sim 0.0$ \\ \hline
\hline
\textbf{Cross section after cuts (fb)}	& 0.5 & 1.5 & 0.01 & $\sim$ 0.0\\\hline\hline
$S/\sqrt{S+B}$ @ 100 fb$^{-1}$		& 	3.5			\\\cline{1-2}
$S/\sqrt{S+B}$ @ 500 fb$^{-1}$		& 	7.9			\\\cline{1-2}
\end{tabular}
\caption{ Cut flow of the efficiencies for signal and backgrounds at the 14 TeV LHC in SRII for $M_{H^\pm}=1000$ TeV (BP9). \label{tab:cut_eff_II_9}}
\end{center}
\end{table}

\subsection{Multivariate Analysis}

We can improve the signal-to-background ratio if we are able to utilize all possible discriminating features in the kinematical distribution profiles of signal and backgrounds through the use of multivariate techniques. For this purpose, we utilize the {\tt TOOLKIT FOR MULTIVARIATE DATA ANALYSIS WITH ROOT} ({\tt TMVA}) \cite{root:tmva} in which various multivariate techniques are implemented in an effective and simple manner. We employ the boosted decision trees (BDTs) analysis to get a better discrimination between signal and backgrounds. This has been shown to perform quickly and effectively for HEP classification problems \cite{Yang2005370}. Other algorithms such as Multilayered Perceptron within TMVA were also considered but deemed to slow for the accuracy of classification provided. One major advantage of using the BDT algorithm is that it can handle a large number of input kinematical variables. In general terms, the more variables are included in the input, the better is the signal and background separation. One can construct several kinematical variables which have some discriminatory power to segregate the signal and background events. However, too many variables might reduce the boosting performance. Thus it is crucial to select the most useful variables, which show reasonable potential for discrimination, so as to maximize the boosting performance. In this regard, we include all kinematical variables displayed in the previous section. 

We first train BDT with $5\times 10^{5}$ signal and $10^6$ background events. Then we perform the testing with the number of events normalized by the integrated luminosity. The BDT algorithm used an 850 tree ensemble (``forest'') that required a minimum of 2.5\% of training events to be passed through each tree and a maximum tree depth of 3. Before passing the events to the BDT for multivariate analysis, we apply preselection cuts in order to separate the events in two different signal regions SRI and SRII. In the following, we list the preselection cuts for each of the signal regions SRI and SRII:

{\bf Preselection for SRI}: Events must have one charged lepton $\ell^\pm$, one fat jet ($J_1$) tagged as the Higgs jet and 3 narrow jets with following requirements on $p_T$ and pseudorapidity:
\begin{subequations}
\begin{eqnarray}
p_T^\ell>20~\text{GeV},\quad |\eta_\ell|<2.5  \\
p_T^{J_1}>200~\text{GeV},\quad |\eta^{J_1}|<2.5\\
p_T^{j}>20~\text{GeV},\quad |\eta^{j}|<2.5
\end{eqnarray}
\end{subequations}
In addition, we further require the transverse momentum of $W_{\text{lep}}$ to be greater than 150 GeV, 200 GeV and 250 GeV respectively for the charged Higgs mass of 500 GeV, 750 GeV and 1 TeV.

{\bf Preselection for SRII}: Events must have one charged lepton $\ell^\pm$, two fat jets ($J_1,~J_2$) one of which tagged as the Higgs jet  and one narrow jet with following requirements on $p_T$ and pseudorapidity:
\begin{subequations}
\begin{eqnarray}
p_T^\ell>20~\text{GeV},\quad |\eta_\ell|<2.5  \\
p_T^{J_1}>150~\text{GeV},\quad |\eta^{J_1}|<2.5\\
p_T^{J_2}>100~\text{GeV},\quad |\eta^{J_2}|<2.5\\
p_T^{j}>20~\text{GeV},\quad |\eta^{j}|<2.5
\end{eqnarray}
\end{subequations}

\renewcommand{\arraystretch}{1.1}
\begin{table}[t]
\begin{center}
 \newcolumntype{C}[1]{>{\centering\let\newline\\\arraybackslash\hspace{0pt}}m{#1}} 
\begin{tabular}{ ||C{3.0cm}||C{2.25cm}| C{2.35cm} || C{2.25cm} | C{2.35cm} ||}\hline \hline
&\multicolumn{2}{c||}{\textbf{SRI}}&\multicolumn{2}{c||}{\textbf{SRII}}\\\hline
\textbf{Benchmark Points}&\textbf{Signal Efficiency (\%)} & \textbf{Background Efficiency (\%)}  & \textbf{Signal Efficiency (\%)}  & \textbf{Background Efficiency (\%)}    \\\cline{1-5}
BP1 & 3.2 & 1.6$\times$10$^{-5}$ &4.1 	&3.5$\times 10^{-4}$\\\hline
BP2 & 2.3 & 1.6$\times$10$^{-5}$ &3.6 	&3.5$\times 10^{-4}$\\\hline
BP3 & 1.5 & 1.6$\times$10$^{-5}$ &3.4 	&3.5$\times 10^{-4}$\\\hline
BP4 & 6.4 & 8.1$\times$10$^{-6}$ &4.6 	&1.2$\times 10^{-4}$\\\hline
BP5 & 5.7 & 8.1$\times$10$^{-6}$ &4.5 	&1.2$\times 10^{-4}$\\\hline
BP6 & 4.4 & 8.1$\times$10$^{-6}$ &4.1 	&1.2$\times 10^{-4}$\\\hline
BP7 & 7.9 & 4.0$\times$10$^{-6}$ &4.9 	&4.5$\times 10^{-5}$\\\hline
BP8 & 7.6 & 4.0$\times$10$^{-6}$ &4.8 	&4.5$\times 10^{-5}$\\\hline
BP9 & 6.8 & 4.0$\times$10$^{-6}$ &4.7 	&4.5$\times 10^{-5}$\\\hline
\hline
\end{tabular}
\caption{ Preselection efficiencies for various signal benchmark points and background in signal regions SRI and SRII at the 14 TeV LHC. \label{tab:presel_eff}}
\end{center}
\end{table}

\begin{figure}[h!]\centering
\includegraphics[width=0.45\textwidth]{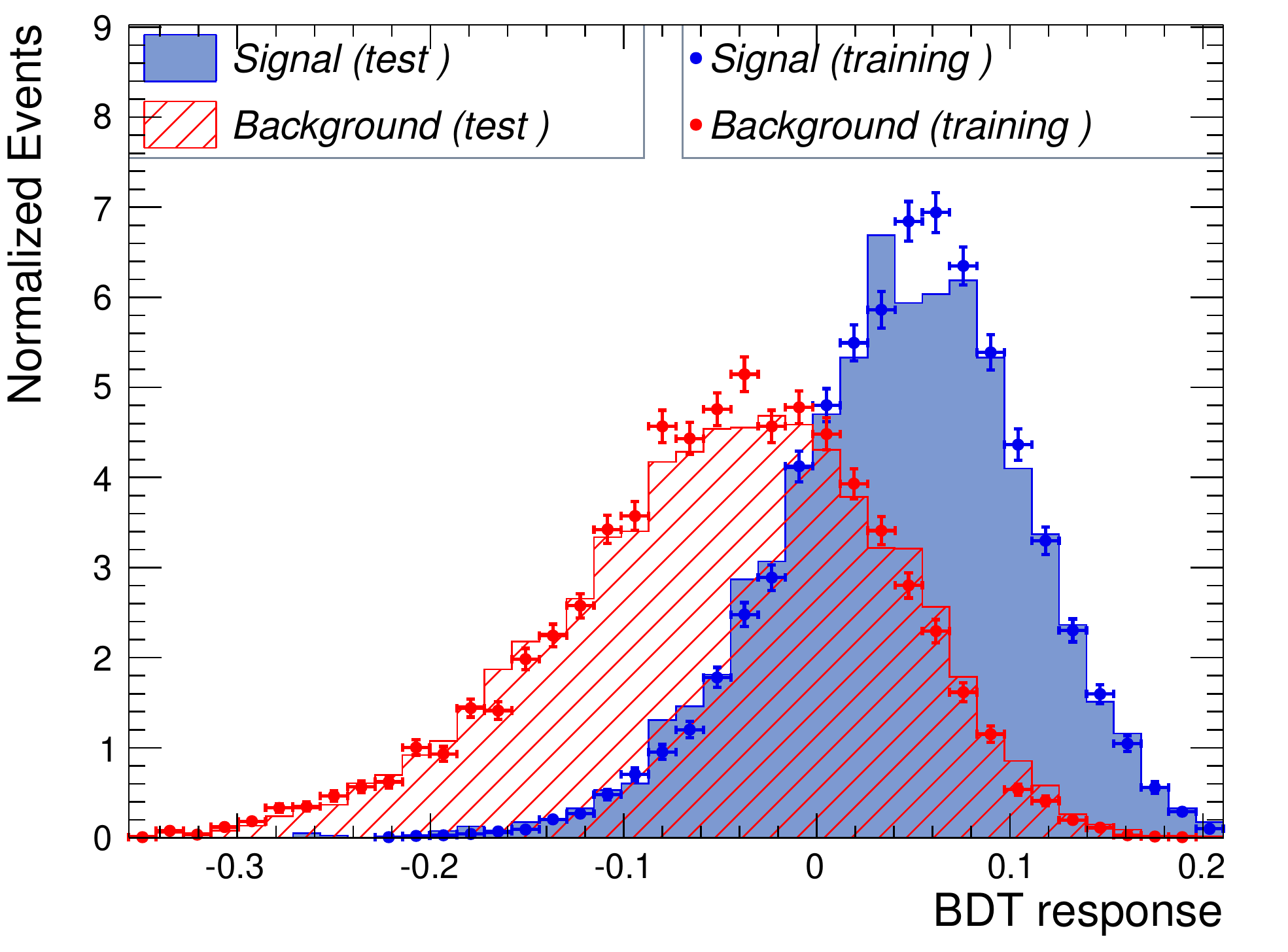}
\includegraphics[width=0.45\textwidth]{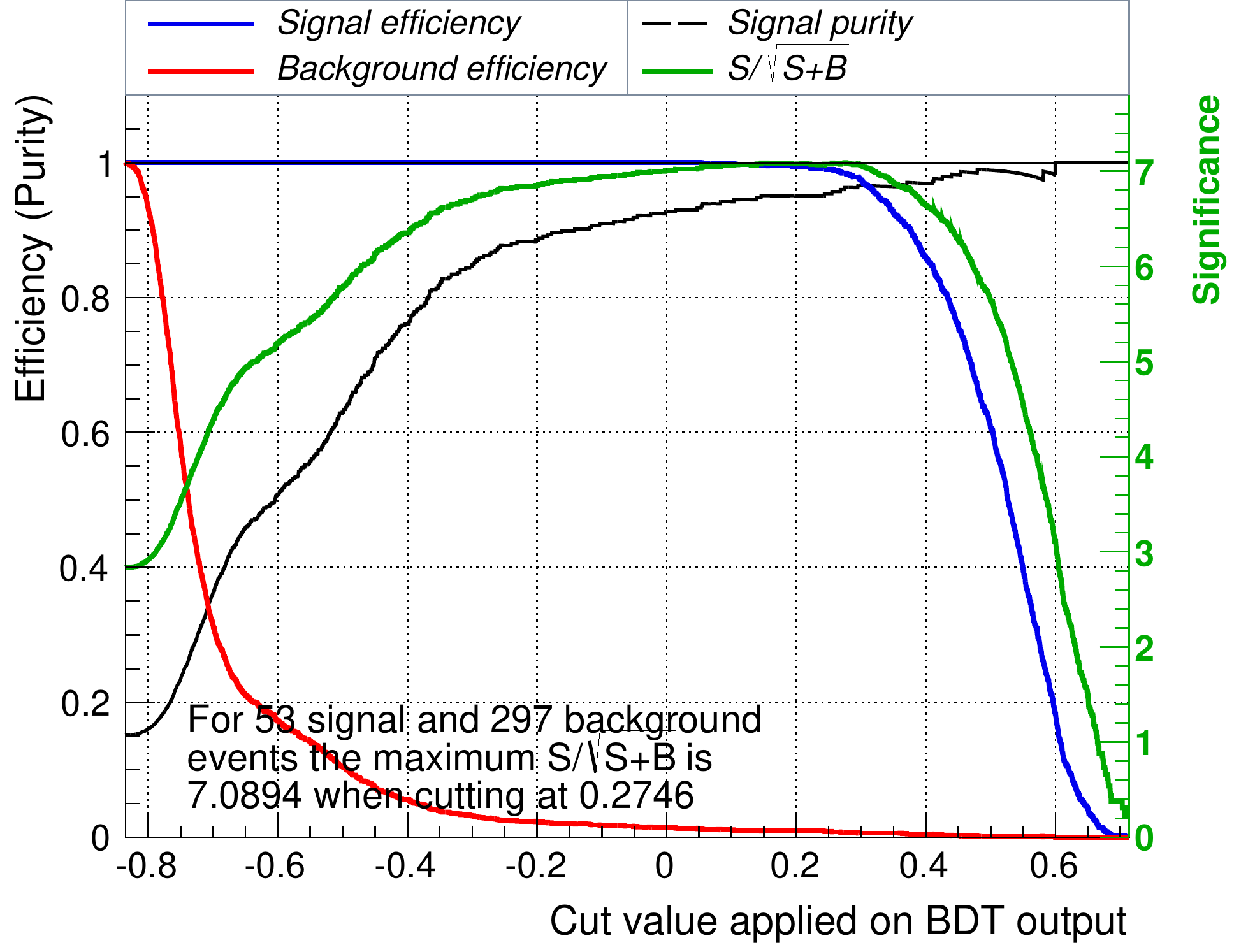}\\\vskip15pt
\includegraphics[width=0.45\textwidth]{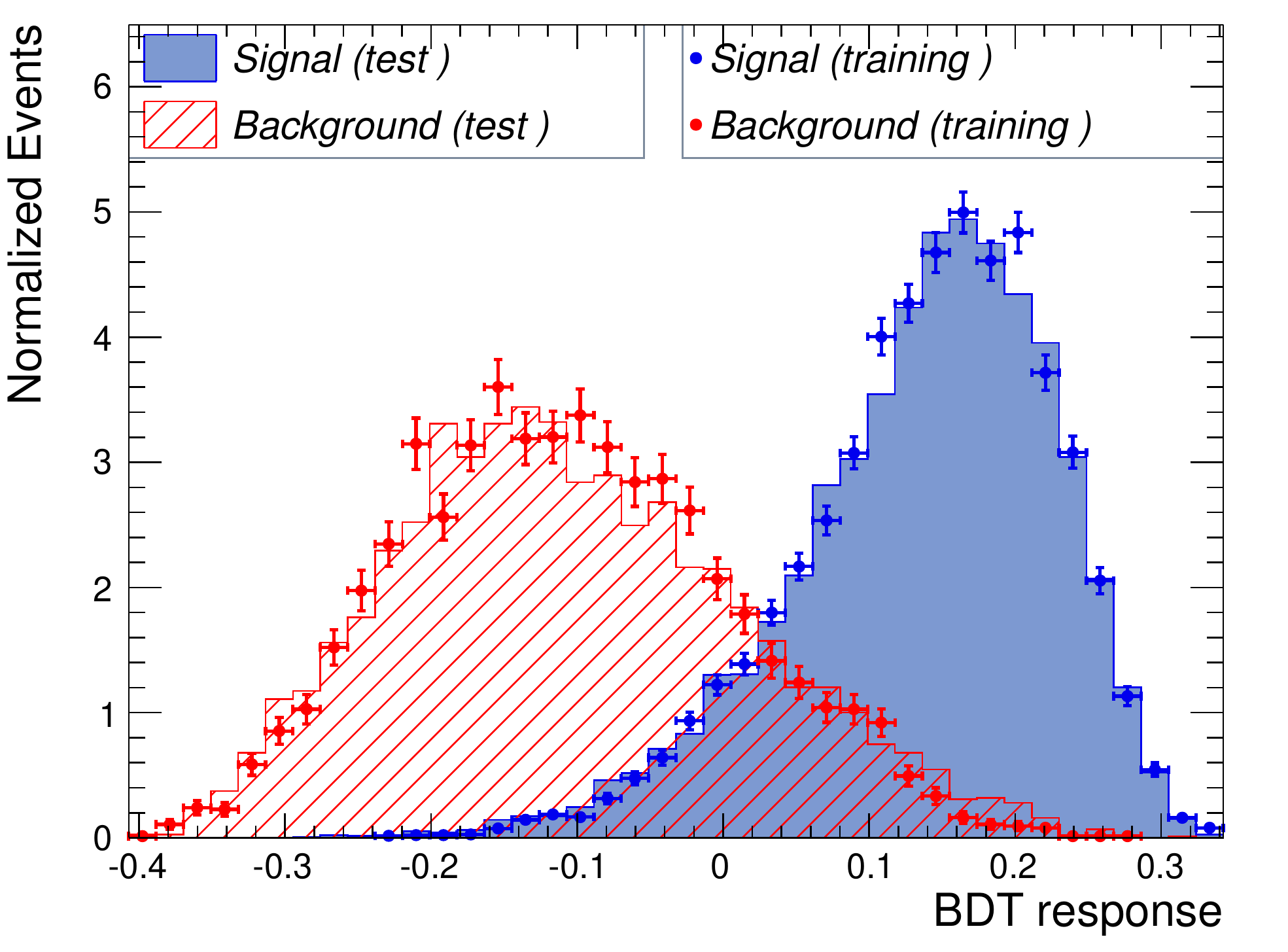}
\includegraphics[width=0.45\textwidth]{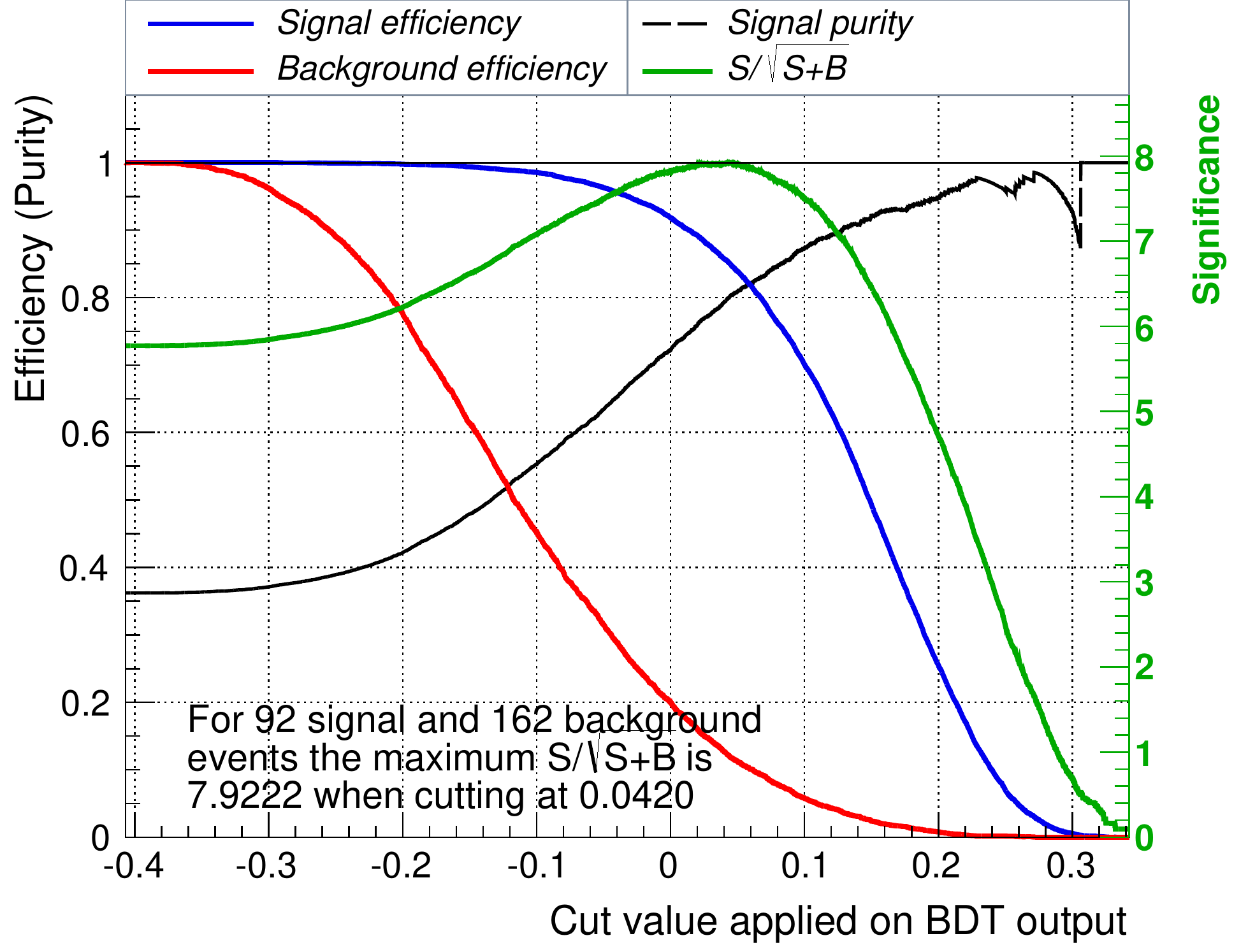}\\\vskip15pt
\includegraphics[width=0.45\textwidth]{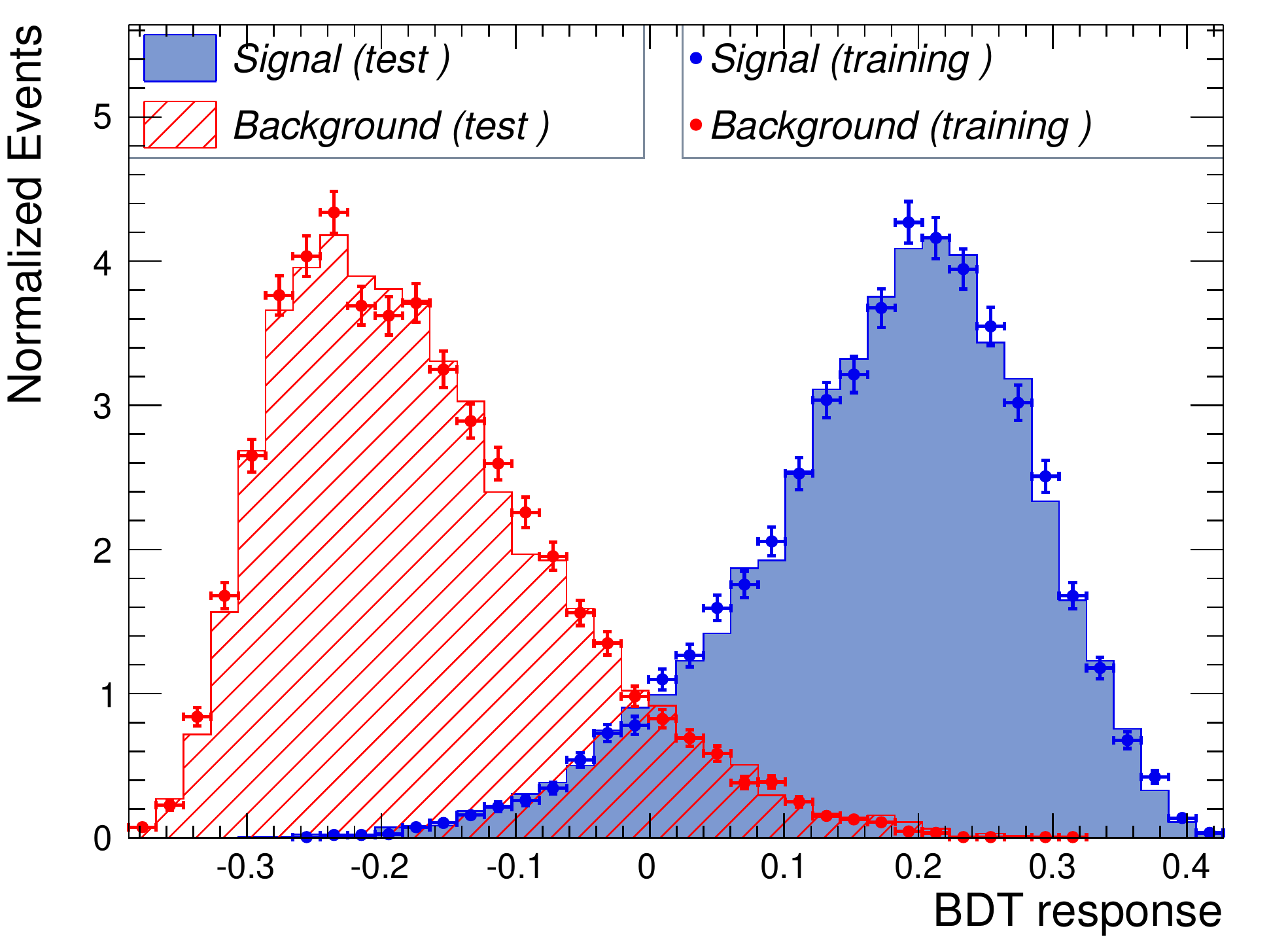}
\includegraphics[width=0.45\textwidth]{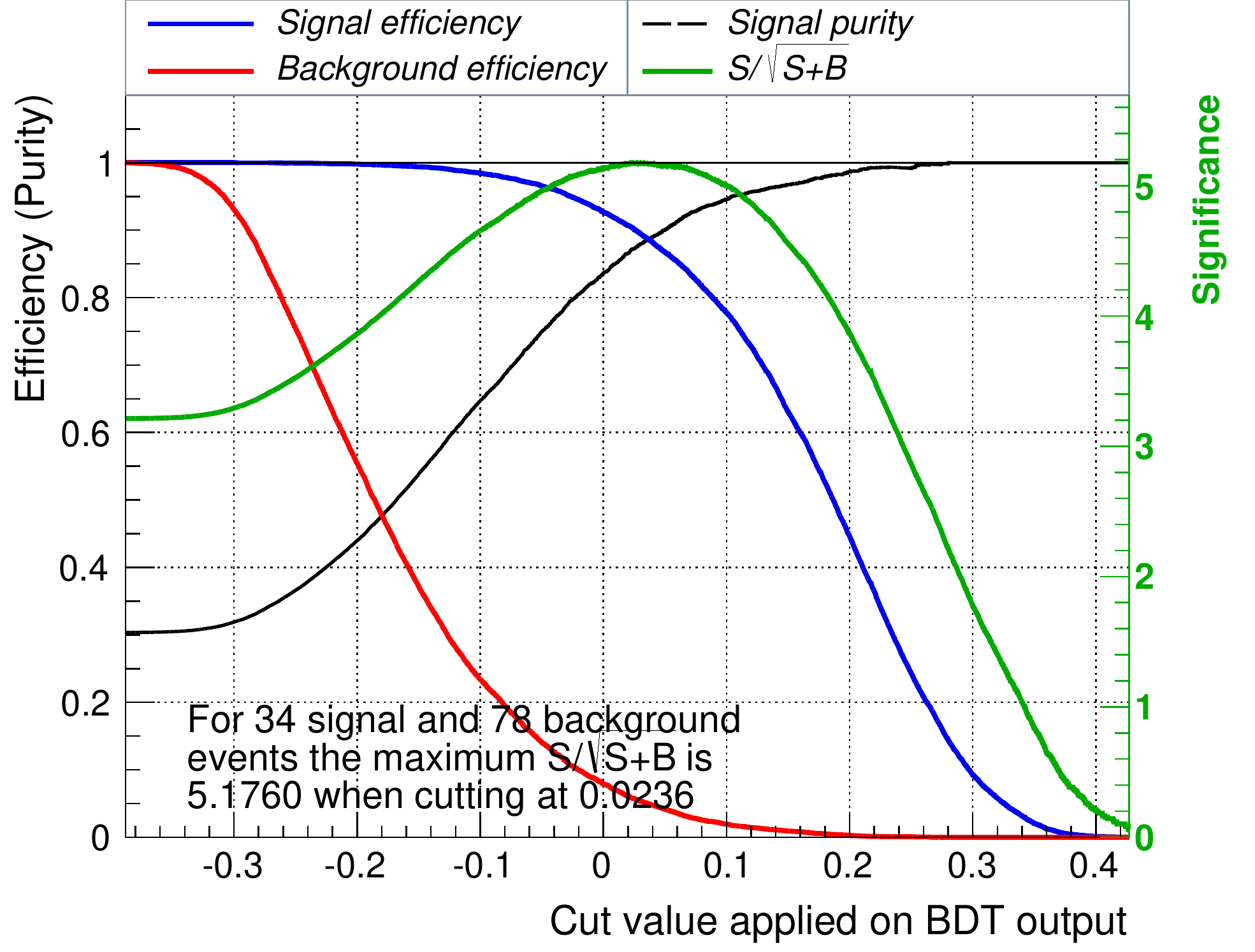}\\\vskip15pt
\caption{\label{fig:BDT_Eff_I}Left: BDT distributions of the signal and background. Right: Signal and background efficiencies as well as signal purity and statistical significance of the signal as a function of BDT cut. The plots are for signal region SRI. The top, middle and bottom panel display the signal benchmark points BP3, BP5 and BP9 respectively. The signal and background test events have been normalized to an integrated luminosity of 100 fb$^{-1}$.}
\end{figure}

\begin{figure}[h!]\centering
\includegraphics[width=0.45\textwidth]{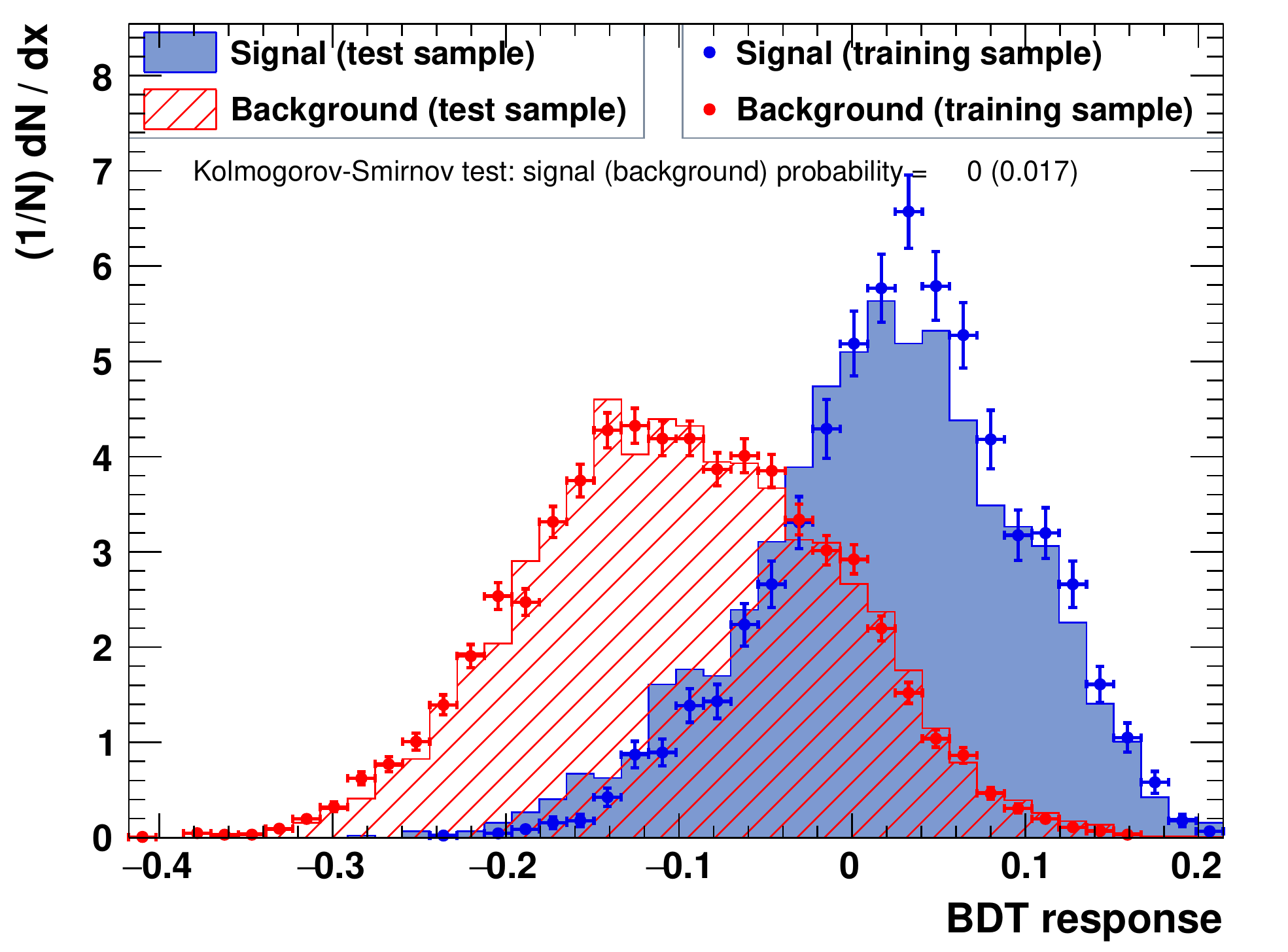}
\includegraphics[width=0.45\textwidth]{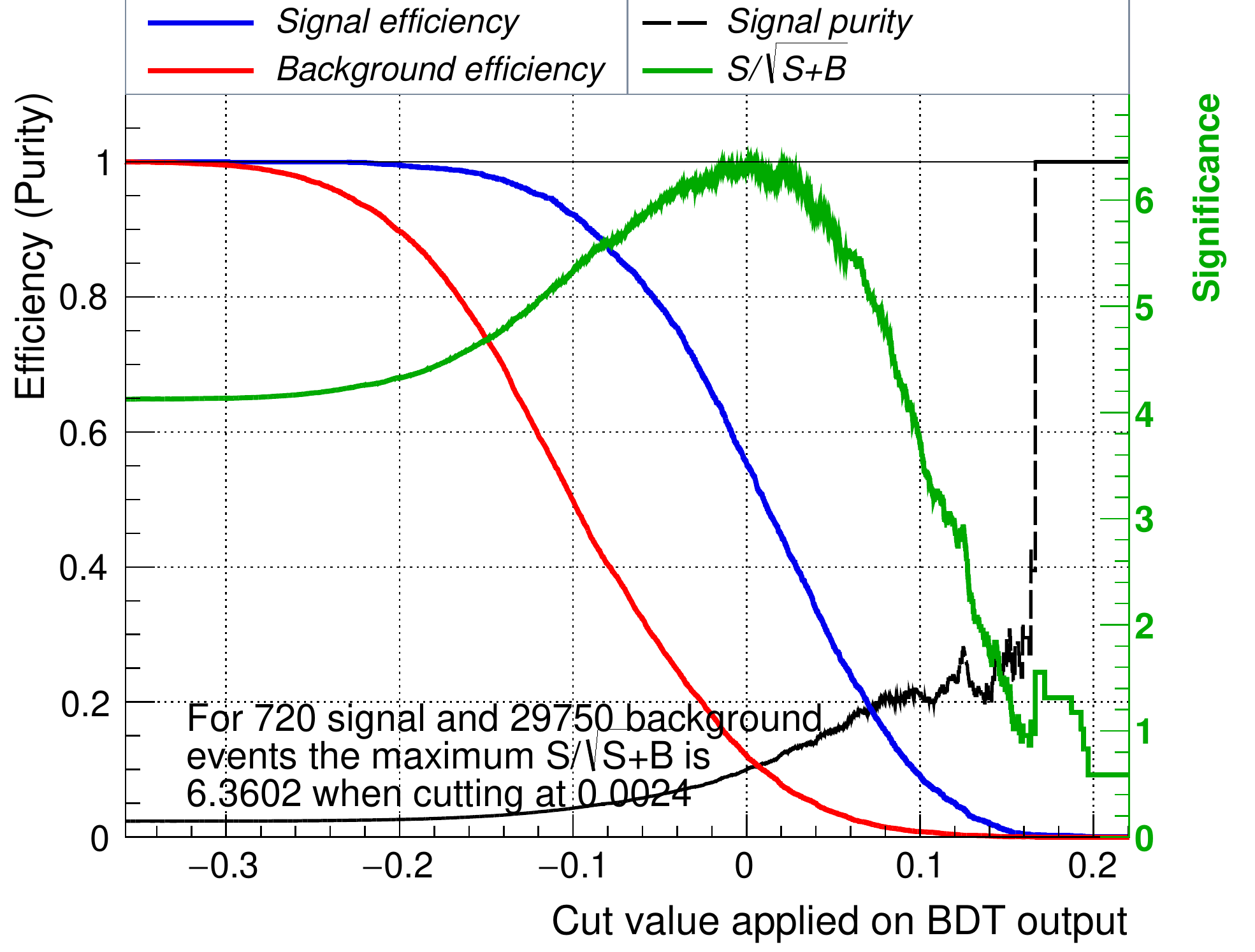}\\\vskip15pt
\includegraphics[width=0.45\textwidth]{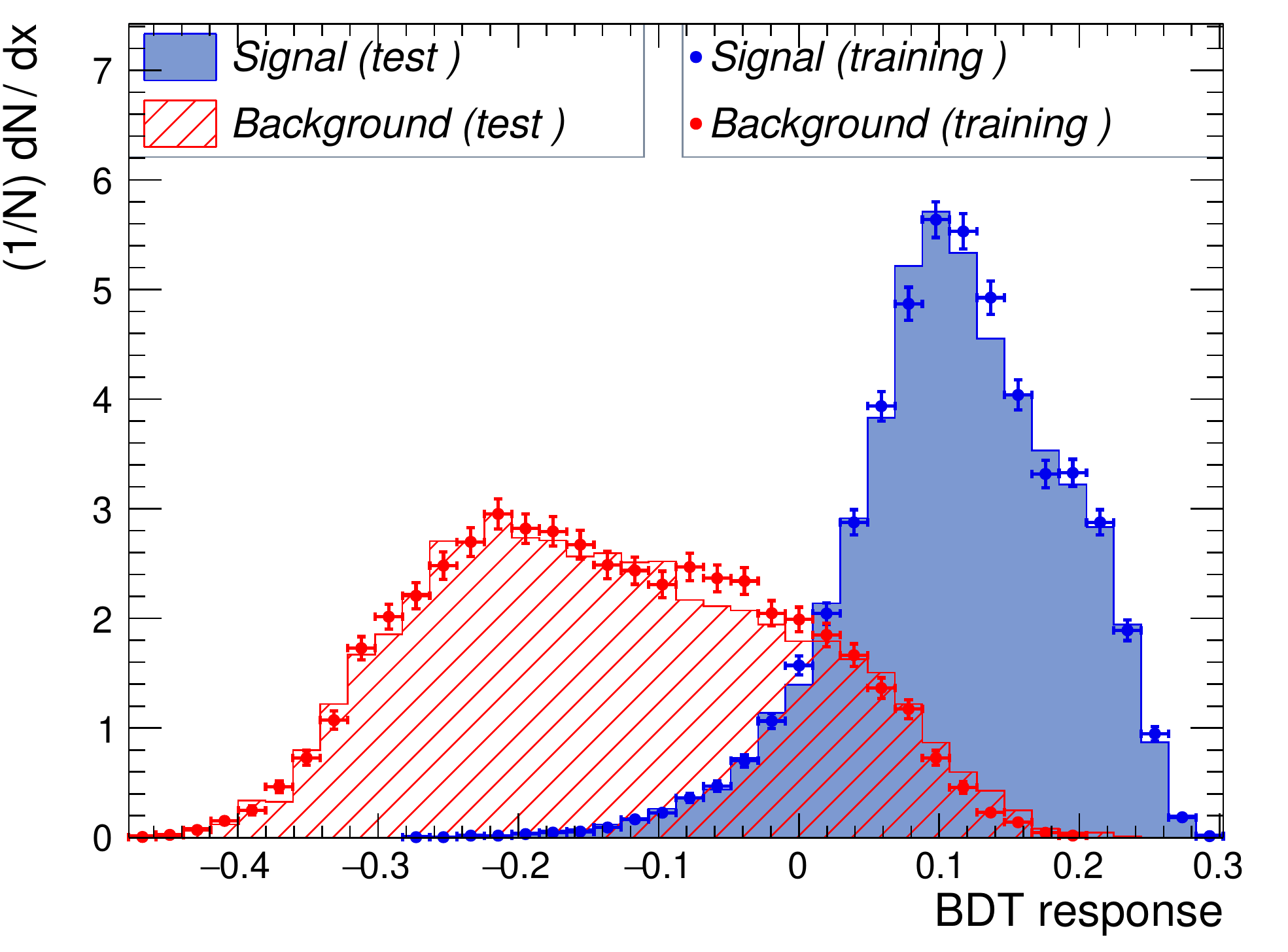}
\includegraphics[width=0.45\textwidth]{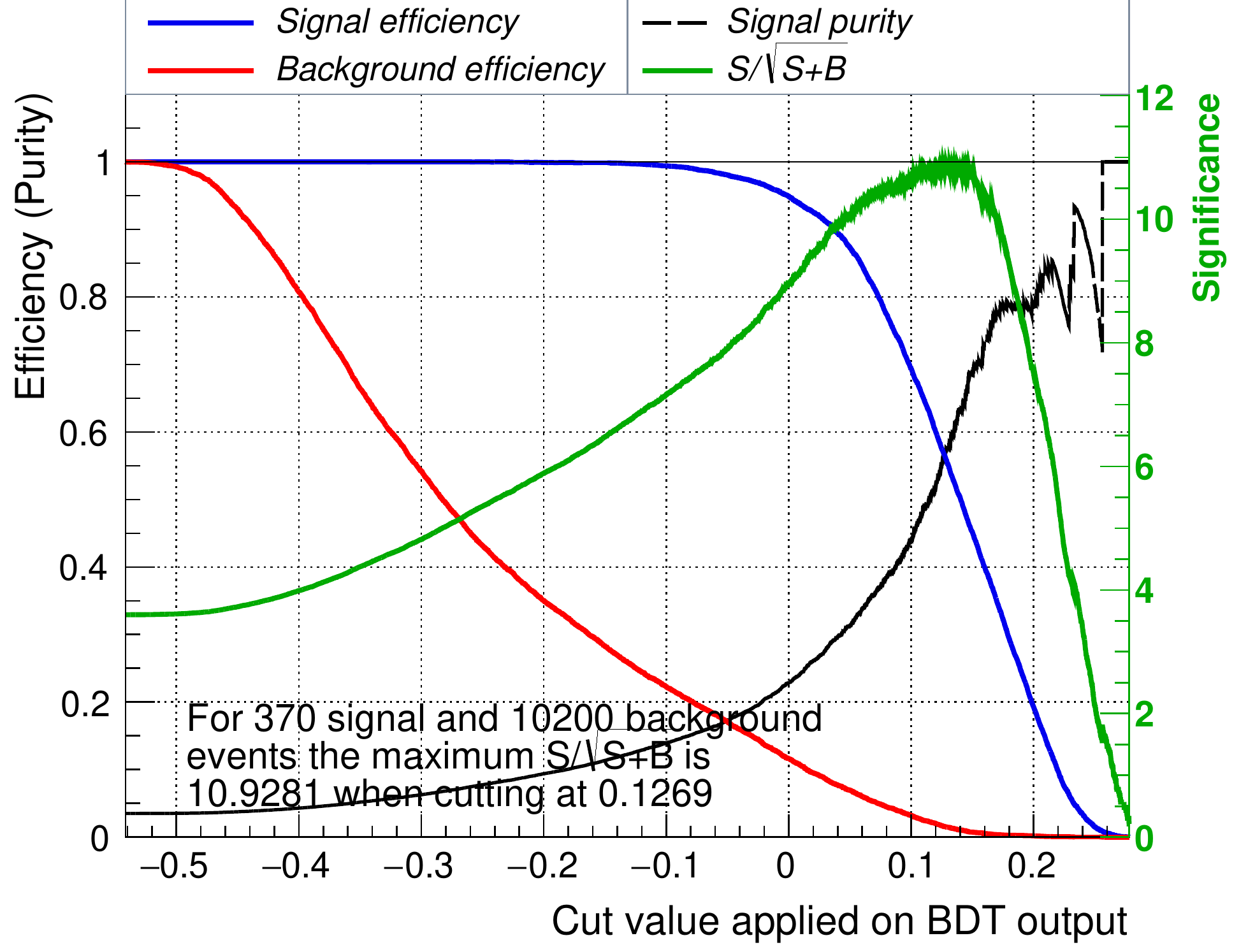}\\\vskip15pt
\includegraphics[width=0.45\textwidth]{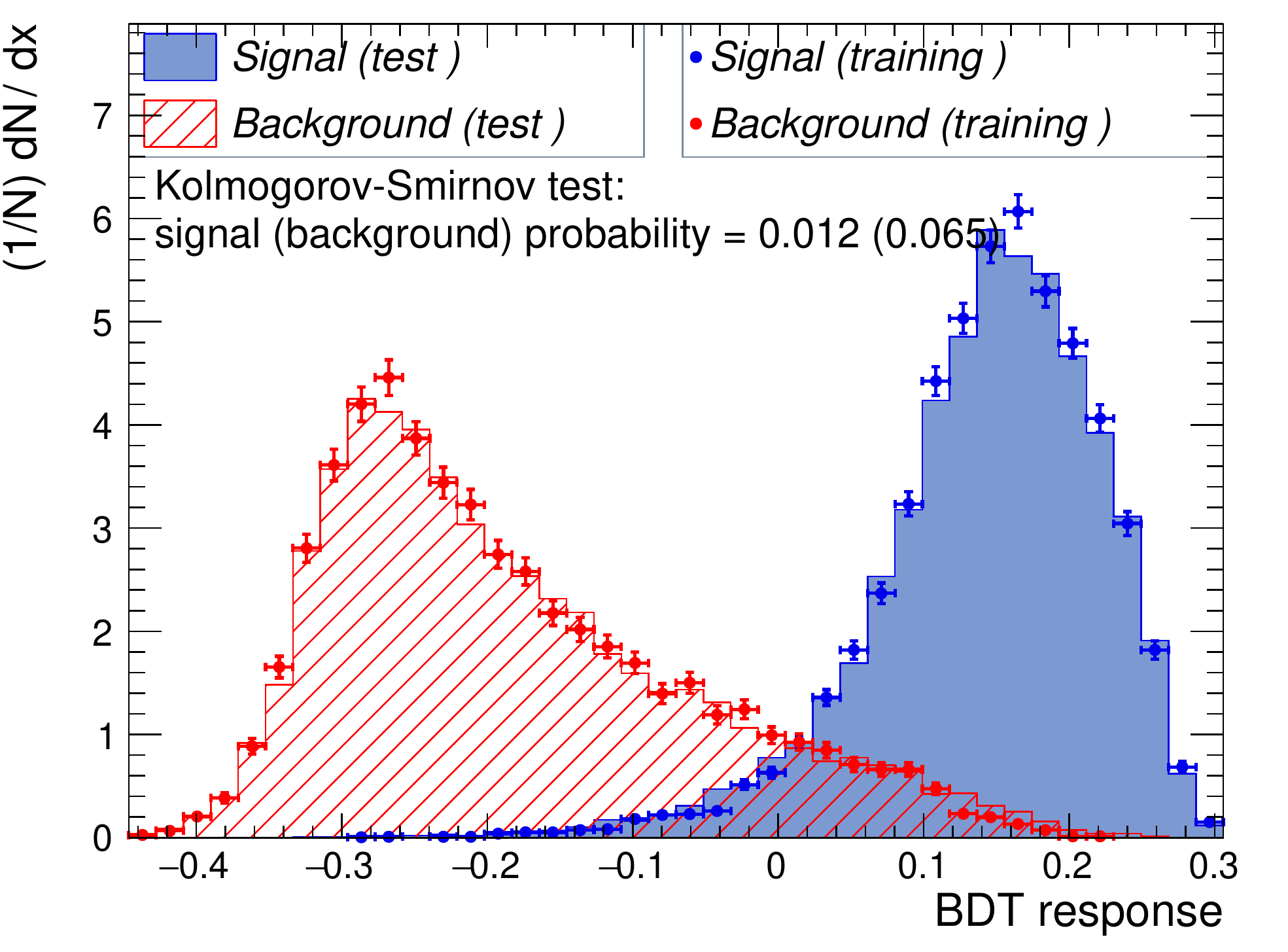}
\includegraphics[width=0.45\textwidth]{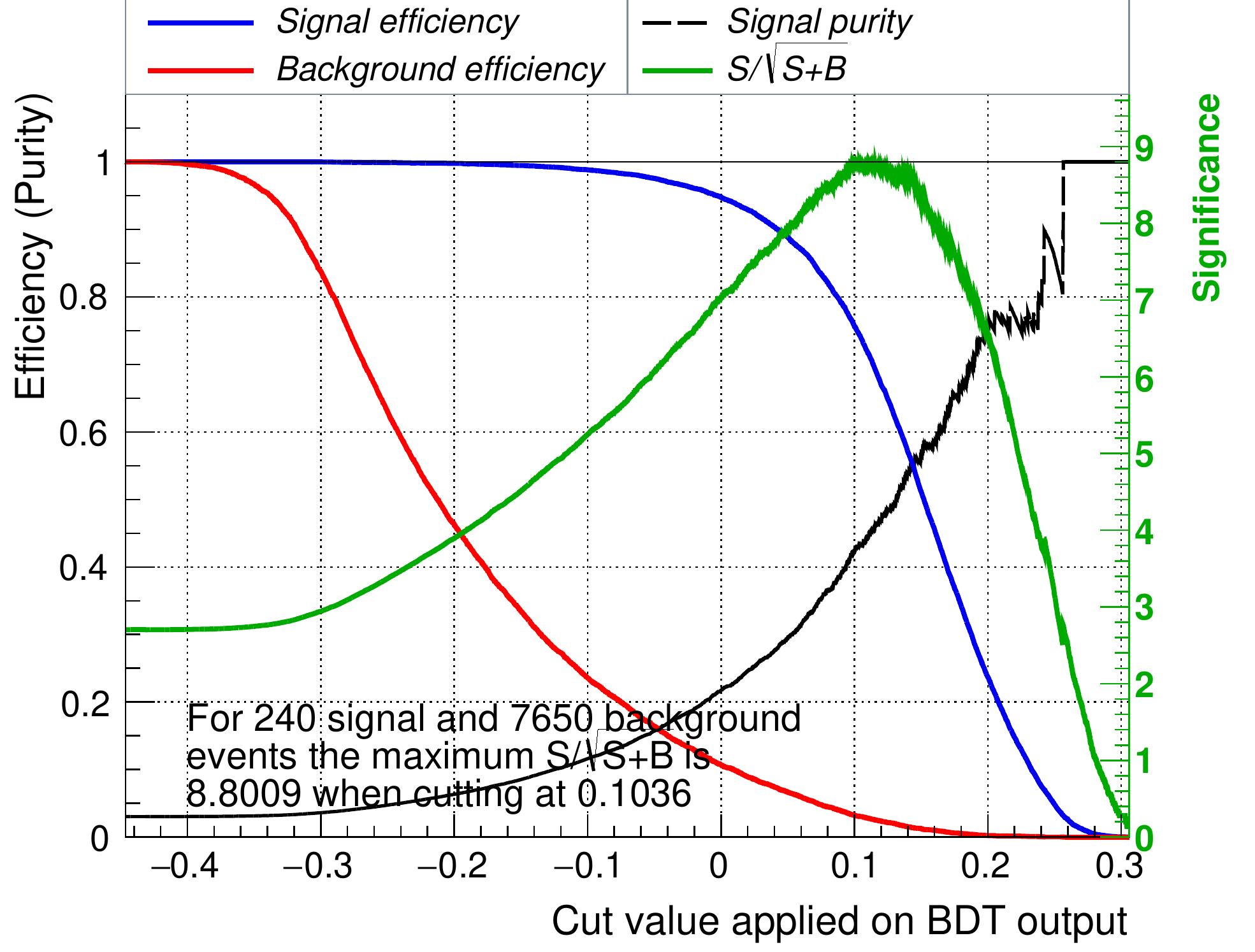}
\caption{\label{fig:BDT_Eff_II} Same as Fig.~\ref{fig:BDT_Eff_I} but for signal region SRII and 500 fb$^{-1}$ of integrated luminosity.}
\end{figure}

The selection cuts are for $M_{H^\pm}$=500 GeV. For a charged Higgs of 750 GeV and 1 TeV, we take $p_T^{J_1}>$ 200 and 250 GeV respectively. Similarly, we slect events with $p_T^{J_2}>$ 150 and 200 GeV for 750 GeV and 1 TeV charged Higgs. In addition, we further require the transverse momentum of $W_{\text{lep}}$ to be smaller than 150 GeV, 200 GeV and 250 GeV respectively for the charged Higgs mass of 500 GeV, 750 GeV and 1 TeV.

The preselection efficiencies for each signal BPs and background in SRI and SRII have been presented in Table~\ref{tab:presel_eff}. In writing the efficiencies we include the $b$ tagging/mistagging efficiencies in the table. For large $H^\pm$ mass, a stringent cut can be applied such that the background can be greatly suppressed while the signal efficiencies are still large. On the other hand, for a low charged Higgs mass of 500 GeV, as the signal and background are less separate, the preselection cuts are not that stringent and lead to large background efficiency. In the signal region SRI, owing to the $p_T$ cut on the leptonic $W$ that is specially introduced to select hard leptonic $W$ bosons from the decay of charged Higgs, the preselection efficiencies for the background events are significantly smaller. Thus in the signal region SRI the background events are expected to be much smaller than in SRII. However in the signal region SRII, the kinematical distribution of variables have far rich features owing to the presence of two hard fat jets. This fact leads to a better performance of MVA in the discrimination of the signal and background events leading to a reasonable signal significance despite having considerably large background events.

\begin{table}[h!]
\begin{center}
\newcolumntype{C}[1]{>{\centering\let\newline\\\arraybackslash\hspace{0pt}}m{#1}} 
\begin{tabular}{ ||C{2.25cm}||C{1.5cm}| C{1.5cm} | C{1.5cm} || C{1.5cm} | C{1.5cm} | C{1.5cm} ||}\hline \hline
&\multicolumn{3}{c||}{\textbf{Signal Significance (SRI)}}&\multicolumn{3}{c||}{\textbf{Signal Significance (SRII)}}\\\hline
\textbf{Benchmark Points}&{100 fb$^{-1}$} & {500 fb$^{-1}$} & {1 ab$^{-1}$} & {100 fb$^{-1}$} & {500 fb$^{-1}$}  & {1 ab$^{-1}$}    \\\cline{1-7}
BP1 & 9.6 & 21.4 & 30.3	& 3.4	& 7.5	& 10.7 \\\hline
BP2 & 7.9 & 17.6 & 24.9	& 3.1	& 7.0	& 9.8 \\\hline
BP3 & 7.1 & 15.8 & 22.3	& 2.8	& 6.4	& 9.0 \\\hline
BP4 & 9.9 & 22.0 & 31.1	& 5.6	& 12.4	& 17.6\\\hline
BP5 & 7.9 & 17.7 & 25.0	& 4.9	& 10.9	& 15.5 \\\hline
BP6 & 8.1 & 18.1 & 25.5	& 4.2	& 9.5	& 13.2 \\\hline
BP7 & 5.7 & 12.8 & 18.1	& 3.8	& 8.5	& 12.1\\\hline
BP8 & 6.0 & 13.5 & 19.1	& 3.5	& 7.9	& 11.1 \\\hline
BP9 & 5.2 & 11.6 & 16.4	& 2.8	& 6.2	& 8.8 \\\hline
\hline
\end{tabular}
\caption{ Signal significance defined as $S/\sqrt{S+B}$ for various signal benchmark points in signal regions SRI and SRII for three different values of integrated luminosities at the 14 TeV LHC. \label{tab:sigtable}}
\end{center}
\end{table}

In the left panels of Figs. \ref{fig:BDT_Eff_I} and \ref{fig:BDT_Eff_II}, we show the BDT distributions for signal and background while in the right panels, the variations of the signal and background efficiencies along with the signal purity and significance have been displayed. The top panel corresponds to benchmark point BP3, the middle panel to BP5 and the bottom panel to BP9 of the signal. The signal purity is defined as the ratio of the signal and the sum of signal and background cross section, $S/(S+B)$, while the signal significance is defined as $S/\sqrt{S+B}$. The plots in the right panel of Figs. \ref{fig:BDT_Eff_I} and \ref{fig:BDT_Eff_II} have been normalized to integrate luminosity of 100 and 500 fb$^{-1}$ respectively. It is evident from the figures that the background efficiency represented by the red curve falls more sharply for 1 TeV $H^\pm$ than for 500 GeV as the far greater mass splitting between the pseudo-scalar Higgs and the charged Higgs is better for substructure analysis.

In Table~\ref{tab:sigtable}, we present the statistical significance of the various signal benchmark points in the two signal regions SRI and SRII for three different chosen values of integrated luminosities, 100 fb$^{-1}$, 500 fb$^{-1}$ and 1 ab$^{-1}$. We find that utilizing a multivariate technique like BDT can significantly enhance the discovery prospects of a charged Higgs. Even with the a integrated luminosity of 100 fb$^{-1}$, the significance of the signal is larger than 5 in most cases. Despite having very small cross section for a 1 TeV charged Higgs, the detection prospects are comparable to the lower $H^\pm$ masses thanks to its suitability for jet substructure analysis.

\section{Conclusions}

The discovery of 125 GeV Higgs-like particle has ushered in a new era in exploration of electroweak symmetry breaking (EWSB) at the large hadron collider (LHC). Various extensions of the standard model EWSB sector introduce additional scalars into the theory. Any further discovery of scalars would be an unambiguous signal of beyond standard model physics. In particular, discovery of a charged Higgs would be a confirmation of an extended scalar sector of EWSB. However, even if a charged Higgs is being produced profusely at the LHC, the search for it is quite complicated due to a large background to its dominant decay to $tb$.

In this work, we study bosonic decays of the charged Higgs that are dominant in certain regions of the parameter space of the two Higgs double models (2HDM). More specifically, we focus on the $H^\pm\to W^\pm A$ decay mode with subsequent decay of the pseudoscalar to a pair of $b$ quarks in associated production of charged Higgs with a top quark. As a charged Higgs of mass lighter than 480 GeV is already ruled out from $b\to s\gamma$ constraints in type II 2HDM, we consider a heavy charged Higgs and consider three different values of its mass 500 GeV, 750 GeV and 1 TeV. We further consider three different masses of the pseudoscalar (100 GeV, 150 GeV and 200 GeV) for each charged Higgs mass. This choice of mass spectrum leads to highly boosted pseudoscalar Higgs in the final state emanating from a heavy charged Higgs decay.  

To enhance the discovery prospects of a charged Higgs in the heavy mass regime, we employ the techniques of jet substructure analysis which play a significant role in tagging a highly boosted Higgs boson. We perform a detailed detector simulation on the 9 signal benchmark points as well as background processes and devise a set of well optimized cuts in a simple cut-based analysis to maximize the signal-to-background ratio. In doing so, we define two signal regions so as to capture the features for in each of the signal regions. The conclusion of the simple cut-based analysis is that a heavy charged Higgs of 1 TeV is discoverable at the 14 TeV LHC with a large significance, while for a 500 GeV $H^\pm$ the signal significance can barely reach 5$\sigma$ even with 3000 fb$^{-1}$ of integrated luminosity.

Finally we perform a multivariate analysis incorporating various kinematical variables that have large discriminating power between signal and backgrounds. We engage the boosted decision tree technique in order to enhance classification performance. We conclude from the MVA that the different distribution profiles of the input variables for the signal and background lead to a very high signal efficiency with respect to background. We find that the charged Higgs would be discoverable with only 100 fb$^{-1}$ of data in the heavy mass region.

\acknowledgments
This work is supported by the University of Adelaide and the Australian Research Council through the ARC Center of Excellence for Particle Physics (CoEPP) at the Terascale (grant no. CE110001004).

\end{document}